\documentstyle[preprint,eqsecnum,aps]{revtex}

\begin{document}
\author{Emil Mottola}
\preprint{LA-UR-95-80}
\address{Theoretical Division T-8, \\Los Alamos National Laboratory, \\
M. S. B285, Los Alamos, NM 87545, USA \\
Electronic Mail: EMIL@PION.LANL.GOV}
\vskip 1truecm
\date{Contribution to the special issue of the\\
Journal of Mathematical Physics\\
on Functional Integration, \\
July, 1995}
\title{FUNCTIONAL INTEGRATION \\
OVER GEOMETRIES}
\maketitle

\newcommand{\sq}{\lower.25ex\hbox{\large$\Box$}}
\newcommand{\sqb}{\lower.25ex\hbox{\large$\stackrel
{\mbox{\rule[.6mm]{2.4mm}{.13mm}}}{\Box}$}}

\begin{abstract}

The geometric construction of the functional integral over coset spaces ${\cal
M}/{\cal G}$ is reviewed. The inner product on the cotangent space of
infinitesimal deformations of $\cal M$ defines an invariant distance and volume
form, or functional integration measure on the full configuration space. Then,
by a simple change of coordinates parameterizing the gauge fiber $\cal G$, the
functional measure on the coset space ${\cal M}/{\cal G}$ is deduced. This
change of integration variables leads to a Jacobian which is entirely
equivalent to the Faddeev-Popov determinant of the more traditional gauge fixed
approach in non-abelian gauge theory. If the general construction is applied to
the case where $\cal G$ is the group of coordinate reparametrizations of
spacetime, the continuum functional integral over geometries, {\it i.e.}
metrics modulo coordinate reparameterizations may be defined. The invariant
functional integration measure is used to derive the trace anomaly and
effective action for the conformal part of the metric in two and four
dimensional spacetime. In two dimensions this approach generates the
Polyakov-Liouville action of closed bosonic non-critical string theory. In four
dimensions the corresponding effective action leads to novel conclusions on the
importance of quantum effects in gravity in the far infrared, and in
particular, a dramatic modification of the classical Einstein theory at
cosmological distance scales, signaled first by the quantum instability of
classical de Sitter spacetime. Finite volume scaling relations for the
functional integral of quantum gravity in two and four dimension

s are derived, and comparison with the discretized dynamical triangulation
approach to the integration over geometries are discussed. Outstanding unsolved
problems in both the continuum definition and the simplicial approach to the
functional integral over geometries are highlighted.

\end{abstract}
\section{Introduction}

\subsection{The Problem of Quantum Gravity}

The beginning of the twentieth century witnessed the birth of two
great breakthroughs in theoretical physics, quantum mechanics and
relativity. Each of these carries with it its own distinct conceptual and
mathematical framework. By the middle of the century, special relativity had
been
incorporated successfully into quantum theory with the introduction of
quantum field theory, though not until the technical difficulties
associated with the handling of infinities and renormalization had
been surmounted. The prototype of all renormalizable quantum field
theories is quantum electrodynamics (QED), which is based on a $U(1)$
gauge invariance. The extension of the gauge principle to non-abelian
groups were then to play a pivotal role in the development of
successful renormalizable theories of the electroweak and strong
interactions of particle physics. The standard model based on the
gauge group ${\cal G} = SU(3)\times SU(2)\times U(1)$ is presently in
agreement with all the known data. In parallel to these developments in
particle physics, Wilson's renormalization group approach to critical phenomena
in statistical systems transformed renormalization from an embarassing
technical blemish of field theory to a positive virtue with genuine physical
implications
\cite{Wil}. The important lesson of Wilson's approach is that renormalizable
theories may play a privileged role in physics only because deviations from
their predictions are naturally suppressed by inverse powers of a very large
ultraviolet cutoff scale. The renormalizable gauge theories of the standard
model, so successful for describing physics at intermediate to low energy
scales may be only effective theories in the infrared. Gradually it has become
clear that the functional integral approach offers a simple, yet powerful
unifying approach to both quantization of gauge theories and the
renormalization group ideas of Wilson.

So far left out of the successful (partial) unification of fundamental
forces in the standard model and the parallel development of
renormalization theory is gravitation. Correspondingly, the geometric
approach to physics so favored by Einstein was relegated to a
secondary or even forgotten role in most of these developments. As we
approach the end of the century it is clear that a consistent theory
of quantum gravity in four dimensions must be regarded as the
outstanding unrealized goal of theoretical physics.

The difficulties in constructing such a theory of quantum gravity are
many, both at the conceptual and the technical level. Certainly the
unrenormalizable ultraviolet divergences present in the Einstein
theory are among the most serious of these difficulties. It is clear
that this problem at the ultrashort Planck scale ($\sqrt {\hbar \over Gc^3}
\simeq 10^{-33}$ cm.) requires the introduction of new physics, and cannot be
solved by
reformulating quantum general relativity in a different way. String theories,
which attempt to solve this problem directly by greatly increasing
the number of degrees of freedom at the ultraviolet scale, is certainly
one possibility for this new physics. Unfortunately, despite prodigious
efforts in the last decade, the technical obstacles in constructing a
completely consistent string theory that agrees with known four
dimensional physics (let alone makes unambiguous predictions for
experimentally testing new physics) still seem very great. Yet, this
effort has not been for naught, since at the very least it has brought
to bear a wealth of new mathematical tools on the problem of
quantizing gravity. One legacy of this development that is likely to
survive in any consistent quantum theory with invariance under
coordinate reparameterizations is the central role played by geometric
methods, if only to extract cleanly the physical predictions of the
theory from the artifacts of coordinate choices. That separation
already presents something of an obstacle to the understanding in the
classical Einstein theory. The importance of formulating the much more
difficult problem of quantum gravity in a coordinate invariant
language, thereby avoiding the cumbersome and distracting technical
hurdles inherent in non-geometric approaches should not be
underestimated. Since the path integral is based on the classical
action, which is invariant under reparameterizations of coordinates,
it is the natural approach to quantization of gravity in which general
coordinate invariance can be maintained in its most transparent form.

In the past, path integral methods have suffered something of a bad
reputation, primarily for the apparent difficulty in defining
precisely integration over such a large function space as the space of
all metrics. It seemed impossible to assign any definite meaning to
such a notion without a discretization procedure which would
inherently introduce some choice of coordinates and immediately wipe
out any advantage of the functional integral approach to quantum
gravity. Just such an invariant definition of functional integration
over two-geometries, {\it i.e.} the coset space of two-metrics modulo
diffeomorphisms ${\cal M}/{\cal G}$, and in particular a much clearer
understanding of the functional measure on this space has been one
important by-product of the geometric approach to string theories \cite{Poly}.
Later it was realized that essentially the same geometric construction could be
taken over to the functional integral over four-geometries directly \cite{BBM}.
The main objection to the use of the path integral method in quantum gravity
was thereby removed.

Parallel to these developments in the continuum definition of the path
integral for systems with coordinate invariance there has also been progress in
an
alternative proposal to quantizing geometry, which has been called the
dynamical triangulation approach \cite{DT}. Based on Regge's discretization of
a continuum manifold by means of triangulation into a finite number of regular
simplices, there are are no coordinates (or coordinate invariances) in this
approach since one sums over distinct geometries from the very outset. Besides
being a conceptually simple formulation well-suited to a systematic numerical
study by computer, the main advantage of the dynamical triangulation approach
is that one can bring all of the tools and intuition gleaned from finite
dimensional statistical systems to bear on quantum gravity. Renormalization
group methods could be applied to coordinate invariant systems in a meaningful
way for the first time, and of course, there are no infinities or ill-defined
functional measures in the discrete definition of the path integral. However,
the infinite volume limit of this formulation must be investigated carefully,
and equivalence with

the continuum functional integral has yet to be demonstrated. In two dimensions
the situation seems reasonably well under control, because of the comparison to
exact results in the continuum for (among other things) the finite area scaling
behavior of the path integral over closed
surfaces with fixed topology. The program of extending the dynamical
triangulation approach to four dimensions has only just commenced in
the last few years \cite{Amb,Mig}, and there the situation is far less well
understood. In this regard, comparisons of numerical data with the recent
predictions for the scaling behavior of the functional integral
over four-geometries with fixed topology should prove very
illuminating \cite{AMM}. The idea here is to make firm predictions at the {\it
infrared} fixed point of the effective continuum theory, independently of the
details at the ultrashort Planck scale, in the same way as Wilson's approach to
critical phenomena are independent of the lattice cutoff. The functional
integral over geometries is the natural tool for analyzing the finite volume
scaling and infrared fixed point behavior of quantum gravity, making contact
between the continuum and dynamical triangulation approaches most directly.

The main purpose of this article is to describe the covariant approach
to the functional integral over geometries from a physicist's (quite
nonrigorous) point of view. After introduction to the general idea of the
geometric approach by means of the quantization of the relativistic point
particle and non-abelian gauge theory, several applications of the covariant
path integral are considered in some detail. While the finite temperature
partition function for gauge theories and the conformal anomaly in
gravitational backgrounds are not new results, it is hoped that their
derivation by means of the covariant path integral will convince physicists of
the simplicity and usefulness of the geometric approach to path integral
quantization, while at the same time motivating a few of the more
mathematically inclined readers to attempt to put these elegant but still
heuristic methods on a firmer mathematical footing. For the benefit of the
latter audience special attention will be paid to open issues and unsolved
problems in the continuum definition of the functional integral. The extension
of the same geometric method to the construction of the functional

 integral over the coset space of metrics modulo coordinate diffeomorphisms is
straightforward, and determines the continuum functional integration measure in
the path integral approach to quantum gravity.

In four dimensions this approach leads to novel conclusions about quantum
gravitational effects at large distances which are discussed next. By comparing
the continuum predictions for the finite volume scaling behavior of the
functional integral to the dynamical triangulation approach, I hope to
communicate some sense of the excitement and anticipation at this frontier of
quantum gravity, as the subject begins to emerge from the shadows of confusion
and speculation into much sharper focus. The solution of the continuum
functional measure problem, the development of numerical techniques in the
dynamical triangulation approach, and the appearance of the infrared
renormalization group in the form of finite volume scaling relations have
introduced some badly needed mathematical clarity and physical intuition into a
field with still very sparse prospects of experimental verification. Future
progress will require an even closer and more fruitful collaboration between
mathematical precision and physical intuition.

The invariant path integral formulation of the problem of quantum gravity as a
sum over geometries seems to offer the best promise of a unifying bridge
allowing just such a fertile interplay of apparently disparate ideas and
techniques, not least
because of the prospect of a third ally soon entering the field, in the
form of numerical results.

\subsection{Basics of Path Integral Quantization}

To establish notation and orientation, let us begin by reviewing the
path integral approach to non-relativistic quantum mechanics described
by a Hamiltonian,
\begin{equation}
H = {p^2 \over 2m} + V(x) \ ,
\end{equation}
where $p={\hbar \over i}{\partial \over \partial x}$ in the position
representation. The evolution of the wavefunction with arbitrary
initial condition at time $t_0$ is described succintly by the
time evolution operator $\cal U$,
\begin{equation}
\psi (x; t) = \int_{-\infty}^{\infty}\ dx_0\ {\cal U}(x,x_0;t-t_0) \
\psi(x_0;t_0)\, ,
\end{equation}
where ${\cal U}$ satisfies the Schr\"odinger Eq.,
\begin{equation}
H\, {\cal U} = i {\partial \over \partial t}\, {\cal U}\,
,
\end{equation}
which has the solution,
\begin{equation}
{\cal U}(x,x_0;t) = \langle x\vert \exp(-iHt)\vert x_0\rangle \,
.\label{krn}
\end{equation}
The path integral representation for this quantity is easily derived
by first dividing the total time interval $t$ into $N$ equal
partitions of length $\epsilon$ (with $t_n = n\epsilon$ and $n=0, 1, \dots ,
N$), and then inserting the identity operator in the form of an
integration over complete sets of position eigenstates of the particle
at each of the $N-1$ intermediate times. In this way we find that
\begin{equation}
{\cal U}(x,x_0;t) = \lim_{N\rightarrow \infty} {\prod_{n=1}^N
\Biggr\{ \int \, dx_n \langle x_n\vert\bigl( 1 - i \epsilon
H(t_n)\bigr)\vert x_{n-1}\rangle \Biggl\}} \delta (x-x_N)\, .
\end{equation}
Evaluating each of the $N$ matrix elements by again inserting the
identity in the form of $N$ integrations over conjugate momentum
eigenstates $\vert p_n\rangle $, and using $\langle p\vert x\rangle =
(2\pi)^{-1}\exp(-ipx)$ yields the desired phase space path integral
representation,
\begin{equation}
{\cal U}(x,x_0;t) = \lim_{N\rightarrow \infty}\prod_{n=1}^N \Biggr\{ \int {dx_n
dp_n\over 2\pi} \exp\biggl\{ -i\epsilon\biggl[{p_n^2 \over 2 m} + V(x_n)\biggr]
+ ip_n(x_n-x_{n-1})\biggr\} \Biggl\}\ \delta (x-x_N) .
\label{pspi}
\end{equation}
An important point about this representation that should be stressed
is that the canonical phase space measure appears with no
arbitrariness whatsover, and the limit $N\rightarrow \infty$ exists
for piecewise smooth potentials, so Eq. (\ref{pspi}) is well-defined.

Further, since all of the $p_n$ integrations are Gaussian, they may be
performed explicitly to arrive at the position space representation of
the Feynman path integral,
\begin{equation}
{\cal U}(x,x_0;t) = \lim_{N\rightarrow \infty} {\prod_{n=1}^{N}
\Biggl\{ \int \, {m\, dx_n \over 2\pi i \epsilon}\exp\biggl\{ {im
\bigl(x_n - x_{n-1}\bigr)^2 \over 2 \epsilon}-i \epsilon V(x_n) \biggr\}
\Biggr\}} \delta (x-x_N)\, .\label{posp}
\end{equation}
Since the quantity in the exponent is precisely the time-discretized
form of the classical action functional evaluated on all possible
paths between the initial position $x_0$ and the final position $x$ in
time $t$, Eq. (\ref{posp}) gives a completely well-defined,
unambiguous meaning to the shorthand notation,
\begin{equation}
{\cal U}(x,x_0;t) = \int_{x(0) = x_0}^{x(t) = x} [{\cal D} x]\,
\exp\bigl\{iS[x]\bigr\}\, ,\qquad S[x] = \int_0^t du \biggl[{m
\dot{x}^2 \over 2} - V(x)\biggr]\,.
\end{equation}
If $t$ were replaced by $-i\tau$ this is exactly the Wiener path
integral which is well studied in the mathematical and statistical
mechanics literature \cite{FKW}.

Although the configuration space path integral is well-defined by the
discretization of the time interval above, such a procedure is awkward
to implement, and is not well-suited to relativistically covariant
Lagrangians where time should play no preferred role. The situation is
reminiscient of the Riemann definition of ordinary integration: it is
important that such a definition exists but it is not how integration
is most easily performed in practice. Instead of the Riemannian
partition of the interval of integration and an awkward limiting
process, it is much more convenient to invoke the fundamental theorem
of calculus and integrate functions by finding their anti-derivative,
completely bypassing any discretization or limiting procedure. It is
at this point that one already encounters difficulties in path
integration, since no Lebesque theory of functional integration is
known, and even the conditions under which a ``fundamental theorem of
functional calculus" should hold are uncertain. In the present state
of development of functional integration we are reduced to either
numerical methods which rely on a discretization definition
similar to (\ref{posp}) or to consideration of a few special cases.
Indeed any ``Table of Functional Integrals" would be very short, since
only Gaussian functional integrals may be evaluated in closed form.
However, for this special case it is possible to give an alternate
definition of the continuum integral which generalizes nicely to
covariant field theories, and which we now discuss.

For a harmonic oscillator action functional we may expand about the
classical solution to quadratic order,
\begin{equation}
S_{quad} = {m\over 2} \int_0^t \, du\, \biggl[\Bigl({dx\over
du}\Bigr)^2 -\omega^2 x^2\biggr] = S_{cl}(x,x_0;t) + {m\over
2}\int_0^t \, du\, \xi\, \Delta_{\omega}\, \xi\, ,
\label{soc}
\end{equation}
where $\xi (u) = x(u) - x_{cl} (u)$ and $x_{cl} (u)$ satisfies the
boundary conditions, $x_{cl} (0) = x_0$ and $x_{cl} (t) = x$. The
differential operator $\Delta_{\omega}$ is given by
\begin{equation}
\Delta_{\omega} = -{d^2 \over du^2} -\omega^2 \, .\label{svr}
\end{equation}
Since this operator may be diagonalized by finding a complete set of
eigenfunctions $\xi_n(u)$ on the interval $[0,t]$ satisfying vanishing
boundary conditions at the endpoints, we would like to be able to
define the path integral over $\xi$ in terms of the determinant of
$\Delta_{\omega}$ which is just the product of its eigenvalues,
$\lambda_n$. This product is infinite, and requires a regulator, but
just such a regulator is provided by the $\zeta$-function technique.
Define
\begin{equation}
\zeta(s\vert \Delta_{\omega}) \equiv \sum_n
\biggl({\mu^2\over\lambda_n}\biggr)^s\, ,
\label{zfn}
\end{equation}
where $\mu$ is an arbitrary scale parameter.  Eq. (\ref{zfn}) defines a
meromorphic function of $s$ with simple poles on the negative integers. In
particular, the sum converges for $Re (s) > 1$, and this defines a function
which may analytically continued to $s=0$. Since for a finite dimensional
matrix operator,
\begin{equation}
-{d\zeta\over ds}(s=0\vert \Delta_{\omega})=\sum_n \ln
\biggl({\lambda_n\over\mu^2}\biggr)=\ln\det\biggl({\Delta_{\omega}\over\mu^2}\biggr)
\, ,\label{zdf}
\end{equation}
we may use Eq. (\ref{zdf}) to {\it define} the determinant of a
differential operator with an infinite spectrum. Then the Gaussian
path integral (\ref{posp}) with action (\ref{soc}) is defined (up to
the arbitrary parameter $\mu$) by
\begin{equation}
\int_{x(0) = x_0}^{x(t) = x} [{\cal D} x] \exp\bigl\{iS_{quad}[x]\bigr\} =
\det{}^{- {1 \over 2}} \biggl({\Delta_{\omega}\over\mu^2}\biggr)
\exp\bigl\{iS_{cl}(x,x_0;t)\bigr\}\, .\label{gsn}
\end{equation}
The $\zeta$-function is closely related to the heat kernel $K$ defined
by
\begin{equation}
K(\tau\vert \Delta_{\omega}) \equiv \sum_n e^{-\lambda_n \tau}
\end{equation}
through
\begin{equation}
\zeta(s\vert \Delta_{\omega}) = {\mu^{2s}\over \Gamma (s)} \int_0^{\infty}
d\tau\, \tau^{s-1} K(\tau)\ .
\end{equation}
In general, it is easier to work with the $\zeta$-function when the
differential operator has known eigenvalues, but with the heat kernel
when deriving general properties of the determinant. Notice also that
the definition of the functional determinant both agrees with the
definition for finite dimensional matrices and has inherent in it a
definition of a quadratic inner product on the space of functions,
$\xi(u)$.

To illustrate the practicality of this definition
let us evaluate the $\zeta$- function for the harmonic oscillator. The
complete set of eigenfunctions of $\Delta_{\omega}$ satisfying the
vanishing boundary conditions at the endpoints of the interval $[0,t]$
are easily found,
\begin{equation}
\xi_n(u) = \sin \Bigl( {\pi n u \over t}\Bigr)\qquad n=1, 2, \dots \label{eig}
\end{equation}
with the corresponding eigenvalues of $\Delta_{\omega}$ given by
\begin{equation}
\lambda_n = \biggl({\pi n \over t}\biggr)^2 - \omega^2\,.
\end{equation}
Hence the $\zeta$-function (\ref{zfn}) may be written as
\begin{eqnarray}
\zeta (s\vert \Delta_{\omega}) &=& \sum_{n=1}^{\infty} \Biggl[{\mu^2\over
\bigl({\pi n \over t}\bigr)^2 - \omega^2}\Biggr]^s \nonumber \\
&=&\Bigl({\mu t\over \pi}\Bigr)^{2s} \sum_{n=1}^{\infty}
n^{-2s}\Biggl[1-\biggl({\omega t \over \pi n}\biggr)^2\Biggr]^{-s}\, ,
\end{eqnarray}
provided $\omega t$ is not $\pi n$ for any positive integer $n$. Let us assume
that $\omega t < \pi$ for the sake of simplicity , although generalizations are
straightforward. Using the binomial expansion for the last term in this
expression,
{\it viz.},
\begin{equation}
(1 - x^2)^{-s} = \sum_{k=0}^{\infty} {\Gamma (k+s) \over \Gamma (k+1)
\Gamma (s)} x^{2k}\, ,
\end{equation}
and interchanging the order of the $k$ and $n$ sums, we find that the
sum over $n$ may be performed explicitly in terms of the Riemann zeta
function $\zeta_R$:
\begin{equation}
\zeta(s\vert \Delta_{\omega})= \Bigl({\mu t\over \pi}\Bigr)^{2s}
\sum_{k=0}^{\infty}{\Gamma (k+s) \over \Gamma (k+1) \Gamma (s)} \Bigl({\omega t
\over \pi n}\Bigr)^{2k} \zeta_R (2s + 2k)\, .
\end{equation}
This form is now suitable for analytic continuation towards $s=0$.
Evaluating the first derivative there gives
\begin{equation}
{d \zeta(s\vert \Delta_{\omega}) \over ds}\Bigg\vert_{s=0} = 2 \ln
\Bigl({\mu t\over \pi}\Bigr) \zeta_R (0) + 2 \zeta '_R (0) +
\sum_{k=1}^{\infty}{1 \over k} \Bigl({\omega t \over \pi n}\Bigr)^{2k}
\zeta_R (2k)\, .
\end{equation}
The properties of $\zeta_R$ are well known. In addition to the values
of the function and its derivative at $s=0$, the relationship of
$\zeta_R (2k)$ to the Bernoulli numbers is needed. The remaining sum
is easily evaluated with the simple result:
\begin{equation}
{d \zeta(s\vert \Delta_{\omega}) \over ds}\Bigg\vert_{s=0} =\ln
\biggl({\omega \over 2\mu \sin \omega t}\biggr)\, .
\end{equation}
Hence from the definition (\ref{zdf})
\begin{equation}
\det{}^{- {1 \over 2}}\Biggl({\Delta_{\omega}\over\mu^2}\Biggr) = \exp
\Biggl({1 \over 2} {d \zeta(s\vert \Delta_{\omega}) \over
ds}\bigg\vert_{_{s=0}}\Biggr) = \Biggl({\omega \over 2\mu \sin \omega
t}\Biggr)^{1 \over 2}\, ,\label{fde}
\end{equation}
and the path integral (\ref{gsn}) has been evaluated unambiguously up
to an overall normalization constant, without any partitioning of the
time interval $[0,t]$ or the limiting procedure of the previous
definition (\ref{posp}).

The unknown constant $\mu$ is somewhat
analogous to a constant of integration. It can be determined from the
normalization condition on ${\cal U}$ which follows from Eq.
(\ref{krn}), namely
\begin{equation}
{\cal U}(x, x_0 ; t=0) = \delta (x - x_0)\, .
\end{equation}
Equivalently one can compare the form of the time evolution operator
obtained from Eqs. (\ref{posp}), (\ref{gsn}) and (\ref{fde}) for the
harmonic oscillator with the known result for the free particle with
$\omega = 0$. By either method one finds that $\mu$ must be replaced
by $i\pi \hbar\over m$ in order to give the correct answer. This is
the same result for ${\cal U}$ that is obtained by the more laborious
method of evaluating the discretized path integral expression Eq.
(\ref{posp}) directly.

As a further exercise with the harmonic oscillator path integral, one
can verify that substituting $t= -i\beta$ and $x_0 = x$, and
performing the Gaussian integration over $x$ gives the finite
temperature partition function of the oscillator,
\begin{equation}
Z(\beta) = \int_{-\infty}^{\infty}\, dx\, {\cal U}(x, x; -i\beta) =
{\rm Tr} \, e^{-\beta H} = {1 \over 2 \sinh \bigl({\omega \beta \over
2}\bigr)}\ .\label{prt}
\end{equation}
This can be evaluated also by the $\zeta$-function method directly if
one computes the determinant of $\Delta_{\omega}$ in Eq. (\ref{svr})
with periodic boundary conditions in imaginary time, $t = -i\beta$,
otherwise following the same steps as Eqs. (\ref{eig})-(\ref{fde})
above. In this case the parameter $\mu$ drops out of the final answer
and the familiar result (\ref{prt}) is obtained directly.

The important lesson of these elementary examples is that a
well-defined and practical definition of the continuum path integral
exists, which does not require any discretization procedure, at least
for Gaussian integrands. In this case one may pass from the
Hamiltonian phase space path integral to the configuration space path
integral defined entirely in the continuum by the proper definition of
the functional determinant (\ref{zdf}). This definition is the one
which extends easily to systems with gauge or reparameterization
invariance which is the main subject of this article.

\section{The Covariant Functional Integral for Simple Systems}

\subsection{Relativistic Point Particle}

There are several forms of the action for a relativistic point
particle, all of which lead to the same answer. We shall use the form
in flat Minkowski space,
\begin{equation}
S[x;N] = {1 \over 2} \int_0^1\, du\, \Bigl({\dot x^a \eta_{ab} \dot
x^b \over N} - m^2 N\Bigr)\, .\label{rpp}
\end{equation}
This action possesses the reparameterization gauge invariance, ${\cal
G}$
\begin{eqnarray}
x^a(u) &\rightarrow &x^a (f(u)) ,\nonumber \\
N(u) &\rightarrow &{df \over du}\, N(f(u)) ,
\label{rep}
\end{eqnarray}
for any differentiable function $f(u)$ that leaves the endpoints of
the interval $[0,1]$ fixed, {\it i.e.} provided
\begin{equation}
f(0) = 0\,,\qquad f(1) = 1\, .\label{ept}
\end{equation}
The appearance of one factor of the determinant of the
reparameterization, $u\rightarrow f(u)$ in the transformation of $N$
in Eq. (\ref{rep}) characterizes it as a density of weight one. The
coordinate $x^a$ is a density of weight zero, {\it i.e.} a scalar
under coordinate reparameterizations.

The momenta conjugate to $x^a$ and $N$ are:
\begin{eqnarray}
p_a &\equiv& {\delta S \over \delta \dot x^a}= {1 \over N} \eta_{ab}
\dot x^b , \qquad {\rm and}\\
p_N &\equiv& {\delta S \over \delta \dot N} = 0\, ,
\end{eqnarray}
respectively. Since the momentum conjugate to $N$ vanishes
identically, the Hamiltonian system is a degenerate one with a primary
first class constraint, in Dirac's nomenclature. The secondary first
class constraint following from the primary one $p_N = 0$ is
\begin{equation}
\dot p_N = -{\delta S \over \delta N} = p_a\eta^{ab}p_b + m^2 \equiv {\cal H} =
0
\, .
\end{equation}
Because of this degeneracy, the phase space Hamiltonian path integral
quantization cannot be applied without some modification. The correct
modification is that of Batalin, Fradkin and Vilkovisky
(BFV) \cite{BFV}. Since the Hamiltonian $\cal H$ is quadratic in
momenta, there is a simple coordinate space path integral obtained by
integrating out the momenta in the BFV formalism. Like the
non-relativistic example considered in the last section, the canonical path
integral has a completely unambiguous well-defined measure in phase
space, which defines a configuration space measure after the Gaussian
integration over momenta. However, a careful definition of the measure
obtained by integrating over the momentum variables seems to require
again partitioning of the interval $[0,1]$ and a limiting procedure.
An interesting open question is whether the passage from the BFV
integral in phase space to the reparameterization invariant
configuration space integral below is possible without recourse to
discretization methods.  Rather than entering into that discussion of
the BFV phase space integral let us follow a different route and
construct the configuration space path integral by geometric
reparameterization invariance considerations alone \cite{BBM,BaV,Bot}.

The space of all configurations, ${\cal M}$ labelled by the functions
$(x^a(u), N(u))$ may be treated by methods borrowed from Riemannian
geometry. At every ``point" in the function space, $X^i$ we introduce
the cotangent space, ${\delta \cal M}$ labelled by the basis vectors
$\delta X^i =(\delta x^a(u), \delta N(u))$. In Riemannian geometry one
can introduce a metric on a space by defining a quadratic form, the
line element, that maps $\delta {\cal M} \times \delta {\cal M}$ to
the real numbers,
\begin{equation}
(\delta s)^2 = G_{ij}(X) \delta X^i \delta X^j\, .\label{lin}
\end{equation}
The (infinitesimal) invariant volume measure of integration on the
cotangent space, ${\delta \cal M}$
\begin{equation}
d (\delta {\cal V}) = \sqrt {\det G_{ij}(X)}\ d (\delta X^1) \times
d(\delta X^2) \times \dots
\end{equation}
may be chosen to satisfy the Gaussian normalization condition,
\begin{equation}
\int d (\delta {\cal V}) \exp \Bigl\{ - {\scriptstyle{1 \over 2}} \delta X^i
G_{ij} (X) \delta X^j\Bigr\} = 1\, ,
\end{equation}
and immediately induces an invariant volume measure on the full space
${\cal M}$,
\begin{equation}
d {\cal V} = {\sqrt {\det G_{ij}(X)}} \ d X^1 \times d X^2 \times
\dots\ \label{vol}
\end{equation}

To define the path integral for a system possessing a
reparameterization gauge invariance, we have only to define a
quadratic inner product on the cotangent space $\delta \cal M$. Then
the construction of the invariant measure on the function space of all
configurations proceeds exactly along the lines of (\ref{lin})-(\ref{vol}).
Since the path integral is specified by an invariant action functional and an
invariant integration measure, this procedure preserves the classical
reparameterization invariance under quantization. Then by identifying
configurations which differ only by a reparameterization of coordinates, we may
integrate only over {\it equivalence classes} of coordinates, ${\cal M}/\cal G$
in a manifestly gauge or coordinate invariant way.

Let us illustrate this procedure for the relativistic point particle.
Since both $N$ and $\delta N$ have weight one under (\ref{rep}), and
\begin{equation}
\int_0^1 \, N(u) \, du \equiv \tau \label{pti}
\end{equation}
is invariant under reparameterizations (acting from the right), we can define
an invariant inner product on $\cal M$ by
\begin{eqnarray}
\langle \delta N,\delta N\rangle _{_N} &\equiv& \int_0^1 \biggl( {\delta N
\over N} \biggr)^2 N\, du = \int_0^1  {(\delta N)^2 \over N} \, du\qquad {\rm
and}\\
\langle \delta x,\delta x\rangle _{_N} &\equiv& \int_0^1 \delta x^a G_{ab}
\delta x^b\, du = \int_0^1\, \delta x^a \eta_{ab} \delta x^b\, N\, du \,
,\label{innr}
\end{eqnarray}
since $\eta_{ab}$ is the only Lorentz invariant tensor possible in
flat Minkowski space, up to a multiplicative constant, which may be
absorbed into normalization of the measure by
\begin{eqnarray}
\int [{\cal D}(\delta N)] \exp \Bigl\{- {\textstyle\frac{1}{2}} \langle \delta
N, \delta N\rangle _{_N} \Bigr\} & = & 1 ,\\
\int [{\cal D}(\delta x)] \exp \Bigl\{ -\textstyle{\frac{i}{2}} \langle \delta
x, \delta x\rangle _{_N} \Bigr\} & = &1 .
\end{eqnarray}
A factor of $i$ is inserted into the second Gaussian because the Minkowski
metric is pseudo-Riemannian with one negative eigenvalue in the timelike
direction, so that the volume form in (\ref{vol}) remains real. In Euclidean
signature metrics this factor would be absent. These definitions generate an
invariant functional measure on the full space $\cal M$ by eqs
(\ref{lin})-(\ref{vol}). In order to pull this measure back
to an invariant measure on the quotient space of equivalence classes
${\cal M}/\cal G$, let us parameterize the gauge orbits $\cal G$ by
the set of differentiable functions $f(u)$ satisfying the endpoint
conditions, (\ref{ept}). Then an arbitrary lapse function $N(u)$ may
be written in the form,
\begin{equation}
N(u) = \tau {df \over du}\ ; \qquad f(u) = {1 \over \tau} \int_0^u\,
N(u) \, du\, .\label{coo}
\end{equation}
In other words, the gauge equivalence class ${\cal M}/\cal G$ of all
the functions $N(u)$ is characterized completely by the single
parameter $\tau$ defined in Eq. (\ref{pti}), and the gauge fiber $\cal
G$ is coordinatized by $f(u)$.  Hence the integration measure on the
quotient space is given by
\begin{equation}
{[{\cal D}N] \over [{\cal D}f]}[{\cal D}x] = J(\tau) d\tau\, [{\cal D}x]\,
,\label{jtr}
\end{equation}
where $J$ is the Jacobian of change of variables from $N$ to $(\tau,
f)$. The invariant path integral is then
\begin{equation}
G(x,x_0) = \int J(\tau) d\tau \int_{x(0) = x_0}^{x(1) = x} [{\cal D}x]\,
e^{i S[x;\tau]}\ .\label{ivp}
\end{equation}

Our task now is to determine the Jacobian $J$. To do so write the
cotangent space form of Eq. (\ref{coo}),
\begin{equation}
\delta N = (\delta \tau) {df \over du} + \tau {d(\delta f) \over du}\, ,
\end{equation}
and substitute this into the inner product definition (\ref{innr}).
The cross term between $\delta \tau$ and $\delta f$ vanishes by using
the endpoint conditions (\ref{ept}). Changing variables in the last
term from $u$ to $v\equiv f(u)$ and defining
\begin{equation}
\xi (v) \equiv (\delta f)(u)\Big\vert_{_{u=f^{-1}(v)}}\,,
\end{equation}
we obtain
\begin{equation}
\langle \delta N,\delta N\rangle _{_N} = {(\delta \tau)^2 \over \tau} + \tau
\int_0^1 \, dv\,  \xi(v)\biggl[-{d^2  \over dv^2} \biggr] \xi(v)  ,
\end{equation}
after an integration by parts. Now, it is straightforward to verify
that the quantity $\xi (v)$ is invariant under diffeomorphism group
transformations operating from the right, {\it i.e.},
\begin{equation}
\xi (v) \rightarrow \xi (v)\ ,\qquad f(u) \rightarrow f(\alpha (u))\
,\label{rig}
\end{equation}
but that it transforms as a density of weight $-1$ under the {\it
inverse} diffeomorphism group transformations operating from the left,
{\it i.e.},
\begin{equation}
\xi (w) \rightarrow {1 \over\Bigl({d \beta^{-1} (w)\over dw}\Bigr)}\xi
(\beta^{-1} (w))\ ,\qquad f(u) \rightarrow \beta (f(u)) \equiv w\ .\label{lef}
\end{equation}
The only quadratic form in $\xi$ that is invariant under both of these
two transformations is \cite{Poly}
\begin{equation}
\langle \xi ,\xi\rangle  \equiv \int_0^1\, \tau dv\, \, (\tau\xi)^2 (v)\
,\label{iom}
\end{equation}
provided that $\tau$ remains invariant under the first transformation,
but
\begin{equation}
\tau \rightarrow \tau \biggl({d \beta^{-1} (w)\over dw}\biggr)\ \label{ttr}
\end{equation}
under the second. The geometric meaning of these transformations is
not difficult to visualize. In the case of (\ref{rig}), $N(u)$ as
parameterized by (\ref{coo}) transforms according to Eq. (\ref{rep})
under right multiplication of the fiber coordinate $f \rightarrow
f\circ\alpha $, while the coordinate in the quotient space ${\cal
M}/{\cal G}$, $\tau$ remains invariant. In the second transformation
(\ref{lef}), the base space coordinate $\tau$ transforms according to
(\ref{ttr}), in order to cancel the change under left multiplication,
$f \rightarrow \beta \circ f$ and leave the total $N(u)$ invariant.

With the fully invariant inner product on the cotangent space to $\cal
G$ defined by Eq. (\ref{iom}), we may define the integration measure
on $\cal G$ by the Gaussian normalization condition,
\begin{equation}
\int [{\cal D}\xi] \exp \Bigl\{ -\scriptstyle{\frac{i}{2}} \langle \xi,
\xi\rangle  \Bigr\}  = 1\ .
\end{equation}
Returning then to our problem of evaluating the Jacobian $J$ in Eq.
(\ref{jtr}), we find that
\begin{equation}
\langle \delta N,\delta N\rangle _{_N} = {(\delta \tau)^2 \over \tau} +
{1\over \tau^2} \langle \xi ,\Delta_0 \xi \rangle \ ,
\end{equation}
with $\Delta_0$ given by the same second order operator (with $\omega
=0$) whose determinant we evaluated in the previous section. Therefore
we have immediately,
\begin{eqnarray}
1 &=& \int [{\cal D}(\delta N)] \exp \Bigl\{ -\scriptstyle{\frac{i}{2}}
\langle \delta N, \delta N\rangle _{_N} \Bigr\} \nonumber \\
& = &\int\, d(\delta \tau)\, J(\tau)\, \int [{\cal D}\xi]\, \exp \Bigl\{ -{i
(\delta
\tau)^2 \over 2\tau} - {i\langle \xi ,\Delta_0 \xi \rangle \over
2\tau^2} \Bigr\} \nonumber \\
& = & J\, \Bigl({2\pi \tau\over i}\Bigr)^{\frac{1}{2}}\,
\Bigl[\det\Bigl({\Delta_0\over\tau^2}\Bigr)\Bigr]^{-\frac{1}{2}} \nonumber \\
& = & const. \times J\ ,
\end{eqnarray}
where the last result follows by Eq. (\ref{fde}). Hence the Jacobian
$J$ is a constant independent of $\tau$.

The evaluation of the propagator in Eq. (\ref{ivp}) is now
straightforward. The functional integration over $x^a (u)$ with fixed
enpoint conditions is the same as that for the free non-relativistic
particle, and we are left with only a simple integral over the proper
time $\tau$ to perform:
\begin{eqnarray}
G(x,x_0) &=& J\int_0^{\infty}\, d\tau\, (2\pi i \tau)^{-{D\over 2}} \exp\Bigl\{
{i\over 2\tau} (x-x_0)^a \eta_{ab} (x-x_0)^b  -{i\over 2}\tau m^2 \Bigr\}
\nonumber \\
&=& J\int_0^{\infty}\, d\tau \int\, {d^Dp\over (2\pi)^D}\, \exp\Bigl\{{ip_a
(x-x_0)^a -{i\tau\over 2}\,(p_a\eta^{ab}p_b + m^2)}\Bigr\}
\nonumber \\
 &=& \int\, {d^Dp\over (2\pi)^D}\, {e^{ip\cdot (x-x_0)} \over (p^2 + m^2 -i0)}\
,
\label{tin}
\end{eqnarray}
provided that the constant $J= {i\over 2}$ and an infinitesimal
negative imaginary part is added to $m^2$ to define the $\tau$
integral. With this normalization Eq. (\ref{tin}) is recognized as
just the Feynman propagator for the free relativistic scalar field,
which we have obtained here by the path integral treatment of the
reparameterization invariant first quantized particle action
(\ref{rpp}). As in the non-relativistic case, the arbitrary
multiplicative constant is the analog of a constant of integration
which can only be fixed by the normalization condition on $G$.

\subsection{Non-Abelian Gauge Theory}

The Yang-Mills action,
\begin{equation}
S_{inv} [A] = -{1 \over 4 g^2} \int d^4 x F^i_{\mu \nu} F^{i \mu \nu}
\label{YMaction}
\end{equation}
is invariant under non-abelian gauge transformations, $A_{\mu}^i
\rightarrow B_{\mu}$ with $A$ and $B$ related by
\begin{equation}
A_{\mu}^i = \{ {\cal U}^{-1} ( \theta) B_{\mu} {\cal U} (\theta)\}^i
-i\{ {\cal U}^{-1} (\theta) \partial_{\mu} {\cal U}(\theta) \}^i
\ ,\label{YMcoor}
\end{equation}
and $\cal U$ is an arbitrary element of the gauge group at each
spacetime point. Thus, $\cal U$ generates the class of field
configurations gauge equivalent to $B_{\mu}^i$. For $\cal U$ close to
the identity the infinitesimal form of Eq. (\ref{YMcoor}) is:
\begin{equation}
A_{\mu}^i \rightarrow A_{\mu}^{i (\theta)} \equiv A_{\mu}^i +
(\nabla_{\mu}\theta)^i
\equiv A_{\mu}^i + \partial_{\mu} \theta^i + f^{ijk} A_{\mu}^j \theta^k\ ,
\label{YMgaugetrans}
\end{equation}
where $f^{ijk}$ are the structure constants of some non-abelian gauge
group.

We may regard Eq. (\ref{YMcoor}) as setting up a coordinate system in
the space of all gauge fields $\cal M$ with gauge fiber coordinates on
$\cal G$ specified by ${\cal U}(\theta)$ and base space coordinates in
the quotient space $\cal M /\cal G$ specified by $B_{\mu}^i$,
analogous to (\ref{coo}) of the relativistic point particle example.
Then we consider the only Poincare invariant quadratic inner product
on the cotangent space, namely
\begin{equation}
\langle \delta A, \delta A\rangle  = \int d^4 x \ \delta A_{\mu}^i(x) \
\eta^{\mu \nu}\
\delta A_{\nu}^i (x). \label{YMiprod}
\end{equation}

This natural, invariant quadratic form is completely analogous to the
invariant line element $ds^2$ of finite dimensional Riemmanian
manifolds, and therefore induces a natural, invariant volume form on
the space of gauge fields. In the present case this metric is flat,
being independent of $A_{\mu}^i (x)$. Hence the invariant volume form
is just the product of coordinate differentials. We may fix the
normalization of this functional measure by a Gaussian normalization
condition:
\begin{equation}
\int [{\cal D}\delta A_{\mu}^i] \exp \Bigl(- {\textstyle\frac{i}{2}}\langle
\delta A, \delta A\rangle  \Bigr)
= 1. \label{YMmeas}
\end{equation}

The measure, defined in this way is manifestly Lorentz invariant.
Moreover, it is gauge invariant under (\ref{YMgaugetrans}). This is
because it is translational invariant under $\delta A_{\mu}^i (x)
\rightarrow \delta A_{\mu}^i (x) + v_{\mu}^i (x)$, and invariant under
global gauge rotations.  Let us now write
\begin{equation}
A_{\mu}^i = {\overline A}_{\mu}^i + a_{\mu}^i
\end{equation}
where ${\overline A}_{\mu}^i$ is a fixed background Yang-Mills field
and $a_{\mu}^i$ is the variable of integration. The corresponding
$B_{\mu}^i$ is given by
\begin{equation}
B_{\mu}^i = {\overline B}_{\mu}^i + b_{\mu}^i = {\overline A}_{\mu}^i
+ b_{\mu}^i
\end{equation}
where we have used the arbitrariness in the separation of background
field and and integration variable to set ${\overline B} = {\overline
A}$. Then the path integral
\begin{equation}
Z[\bar A] = (Vol({\cal G}))^{-1} \int [{\cal D} a_{\mu}^i] \exp (i
S_{inv}[\overline A + b]). \label{YMZ}
\end{equation}
may be defined with respect to the measure induced by the definitions
(\ref{YMiprod}) and (\ref{YMmeas}). Here $Vol({\cal G})$ denotes
schematically the (infinite) gauge orbit volume that must be divided
out of $[{\cal D} a_{\mu}^i]$, and $b$ denotes the integration
variable over the equivalence classes of potentials in ${\cal M}/{\cal
G}$. In order to specify the field coordinates completely one still
needs to fix a condition on the $b_{\mu}^{i}$ of the form,
\begin{equation}
(F \cdot b)^i = 0.
\label{YMconstraint}
\end{equation}
For example $F$ might be the covariant derivative with respect to the
background gauge potential $\bar A$, in which case the results will be
formally identical to those obtained in the background field Landau
gauge. However, it is important to emphasize that in the geometric
approach follwed here this is not a gauge ``fixing" that breaks the
gauge invariance of the theory, but rather merely a particular choice
of coordinates in a manifestly coordinate ({\it i.e.} gauge) invariant
construction, since $b_{\mu}^{i}$ lies by definition in the quotient
space ${\cal M}/{\cal G}$.

Again, our task is to express the gauge invariant quantity (\ref{YMZ})
in the coordinates (or gauge) specified by (\ref{YMcoor}) and
(\ref{YMconstraint}).  We first compute the Jacobian of the
transformation to the new coordinates
\begin{equation}
[{\cal D}a_{\mu}^i] = J\, [{\cal D}b_{\mu}^i] [{\cal D}\theta^i],
\end{equation}
in the tangent space:
\begin{eqnarray}
1& = &\int [{\cal D}a_{\mu}^i] \exp (- {\textstyle\frac{i}{2}} \langle
a,a\rangle ) \nonumber \\
& = &\int J\, [{\cal D}b_{\mu}^{i}] [{\cal D}\theta^i] \exp \bigl( -
{\textstyle\frac{i}{2}}
(\langle b, b\rangle + 2 \langle \nabla \theta, b\rangle + \langle \nabla
\theta,
\nabla \theta\rangle ) \bigr)
\end{eqnarray}
This integral may be computed by completing the square and using
condition (\ref{YMconstraint}). We find that
\begin{equation}
J = [\det{}_S (-\nabla^2)\ \det{}_{V}\bigl(\delta_{\mu}^{\nu} -
\nabla_{\mu}(\nabla^2)^{-1} \nabla^{\nu}\bigr)\big\vert_{_F}]^{{1 \over 2}},
\label{YMjacobian}
\end{equation}
where the vector determinant is to be evaluated over the space of
fields $b_{\nu}^{i}$ obeying the condition (\ref{YMconstraint}).  This
vector determinant may be converted into a scalar
determinant \cite{BBM} and the Jacobian (\ref{YMjacobian}) is given
finally by:
\begin{equation}
J = \det{}_S^{- {1 \over 2}} (F\cdot F^{\dagger})\
\det{}_S(F\cdot \nabla) \label{YMjacobianb}
\end{equation}
where $F^{\dagger}$ is the adjoint of $F$ with respect to the inner
product (\ref{YMiprod}). In this form connection to the standard
gauge-fixed perturbative formalulation of Yang-Mills theory may be
made, since the second factor in (\ref{YMjacobianb}) is precisely the
Fadeev-Popov Jacobian $\Delta_{FP}$ of the gauge fixing method for the
gauge condition (\ref{YMconstraint}), while the first factor is a constant,
independent of the field point $A_{\mu}$, and therefore may be taken out of the
path integral, without affecting the result. It is important to retain this
factor if one calculates the effective action as a function of the background
field, however.

Having determined the correct Jacobian of the transformation to field
coordinates, $(b_{\mu}, \theta)$, we may now express $Z$ in the form,
\begin{eqnarray}
Z [\bar A]& = &(Vol({\cal G}))^{-1}\int [{\cal D}\theta^i] \int J\ [{\cal D}
b_{\mu}^{i}] \exp (i S_{inv}[\bar A + b]) \nonumber \\
& = &\det{}_S^{-{1 \over 2}}(- \overline \nabla^2)
\int [{\cal D}b_{\mu}^{i}] \det{}_S(-\overline \nabla \cdot
\nabla)\big\vert_{A=\overline A + b}\exp (i S_{inv} [\overline A + b]),
\label{YMfinal}
\end{eqnarray}
where the backgound Landau gauge $F^{\mu} = \overline\nabla^{\mu}$ has been
chosen and the integral is over the gauge invariant field coordinate,
$b_{\mu} ^i$. The gauge volume factor has been cancelled
explicitly. If one expands the invariant action in a power series in
the integration variable $b$, treating the quadratic Gaussian
functional integral exactly, but the higher powers of $b$ as
perturbations, it is clear that this form generates the same Feynman
rules as the gauge fixed path integral of more standard formulations,
and that a similar equivalence holds for other choices for the gauge
fixing function $F$. In the present approach, the geometric
significance of gauge fixing as simply a choice of coordinates in a
coordinate invariant expression is manifest and proofs of gauge
invariance are unnecessary. Moreover, the non-perturbative aspects of
the path integral and the correct dependence on the background field
$\overline A$ in $\det{}_S^{-{1 \over 2}} (- \overline \nabla^2)$ has
been obtained as well. This background field dependence is essential
for the proof of Ward identities in the background field method, and
may be lost in a careless treatment of the gauge-fixing. The
background field method is an extremely efficient method of extracting
physical results, such as asymptotic freedom in non-abelian gauge
theories, with a minimum of technical complication. Provided the
functional determinants in these formal expressions can be defined by
a suitable invariant regulator, such as the $\zeta$-function method of
the last section, (\ref{YMfinal}) defines the configuration space path
integral for Yang-Mills theory in a manifestly gauge and Lorentz
invariant manner without recourse to discretization and limiting
procedures.

One possible problem that this construction of the functional integral for
non-abelian gauge fields glosses over is the question of a {\it global} choice
of coordinates on the coset space. The construction assumes that we can choose
coordinates on the full configuration space such that the slice specified by
(\ref{YMconstraint}) in the coset space ${\cal M} \over {\cal G}$ intersects
each gauge fiber $\cal G$ once and only once. This is by no means guaranteed in
general, and in fact, it is known that for certain common choices of $F$ this
is {\it not} the case \cite{Gri}. An interesting open issue is whether or not
it is possible to find a global choice of coordinates on the configuration
space that does satisfy this criterion.

\subsection{The Partition Function for a Finite Temperature Photon Gas}

As a simple application of the geometric path integral in the case of
a theory with an internal gauge group, let us consider how one would
calculate the partition function and thermodynamics of a gas of free
photons. If we set the background field $\bar A = 0$ in the expression
(\ref{YMfinal}), and continue to imaginary time, $t \rightarrow it_E,
iS_{inv} \rightarrow -S_{E}$ we obtain the formal result for the
partition function $Z$ for a general Yang-Mills theory with action
(\ref{YMaction}). If we take the gauge group to be the abelian group
$U(1)$ of electromagnetism, so that the structure constants $f_{ijk}
=0$ as well, then $\nabla = \bar \nabla = \partial$ and $Z$ reduces to
\begin{equation}
Z = \det{}_S^{+{1 \over 2}}(- \sq) \int [{\cal D}b_{\mu}^{i}] \exp
(-S_{_E} [b])
\label{QEDint}
\end{equation}
where the Euclidean action may be written in the form,
\begin{equation}
S_{_E}[b] = +{1 \over 4 g^2} \int d^4 x_{_E} F_{\mu \nu} F^{\mu \nu}
= {1 \over 2 g^2} \int d^4 x_{_E} b_{\mu}(-\sq )b_{\mu}
\end{equation}
in the Landau gauge $F\cdot b = \partial_{\mu} b_{\mu} = 0$. Then the
Gaussian path integral (\ref{QEDint}) is immediately performed to
yield
\begin{equation}
Z = \det{}_S^{+{1 \over 2}}(- \sq)\det{}_V^{-{1 \over 2}}(-
\sq)\Big\vert_{F}\ ,
\end{equation}
where the vector determinant is to be evaluated over vectors
satisfying the Landau transversality condition. Since any four vector
may be decomposed into transverse and longitudinal components which
decomposition is orthogonal with respect to the inner product
(\ref{YMiprod}), we may extend the evaluation of the vector
determinant over all four vectors provided we divide out the
longitudinal contribution, {\it i.e.}
\begin{equation}
\det{}_V(- \sq)\Big\vert_{F} = \det{}_V(- \sq) /\det{}_S(- \sq) = \det{}_S^3(-
\sq)
\end{equation}
where the last step follows by introducing a Cartesian basis for the
four vectors, so that the unrestricted vector determinant is just the
determinant of the scalar operator $- \sq$ raised to the fourth power.
Hence,
\begin{equation}
Z = \det{}_S^{-1}(- \sq)\ ,
\label{QEDZ}
\end{equation}
which is the same as the result for two free massless scalar fields.
These are the two physical helicity states of a massless photon (or
gluon). In this form the significance of the Jacobian factor
(\ref{YMjacobianb}) may be understood. Its role is precisely to cancel
unphysical degrees of freedom in the covariant four vector gauge
field, reducing the contribution to just the correct physical degrees
of freedom of the canonical method of quantization.

To complete the evaluation of the partition function for the photon
gas we have only to note that finite temperature implies that the
operator $- \sq$ should be evaluated over functions periodic in
Euclidean time with period $\beta$ equal to the inverse temperature
$T^{-1}$. Hence its eigenvalues are just
\begin{equation}
\biggl({2\pi n \over \beta}\biggr)^2 + \vec k^2
\end{equation}
and the result for $Z$ in (\ref{QEDZ}) is nothing but the square of
the product over wavenumbers $\vec k$ of the harmonic oscillator
result (\ref{prt}) with the frequency $\omega$ replaced by $\omega_k =
\vert \vec k \vert$, {\it i.e.}
\begin{equation}
Z \equiv e^{-\beta F} = \Biggl[\prod_{\vec k} {1\over 2 \sinh \bigl(
{\omega_k \beta \over 2 }\bigr)}\Biggr]^2\ ,
\end{equation}
so that the free energy function is
\begin{equation}
F = -T \ln Z = 2 T\sum_{\vec k}\ln \biggl(e^{{\omega_k \beta \over 2
}} - e^{-{\omega_k \beta \over 2 }}\biggr) = \sum_{\vec k} \omega_k +
2 T \sum_{\vec k} \ln \bigl( 1 - e^{-\omega_k \beta}\bigr)\ ,
\label{QEDF}
\end{equation}
the first term of which is the infinite, but temperature independent
sum of zero point energies $\omega_k \over 2$ for each of the two
photon helicity states. This ultraviolet quartic divergence in the
zero temperature free energy is a general feature of field theory, and
leads to the ``cosmological constant problem" when gravitation is
considered, since there are no observable effects of this zero-point
energy on the curvature of spacetime. This divergence would be fully
regulated by the $\zeta$-function method if we performed the sum over
$\vec k$ (as well as $n$) before analytic continuation to $s=0$,
although this involves a more difficult calculation. In any case, the
question of the absolute zero point of free energy does not arise if
we are interested only in energy differences or the temperature
dependence of the free energy, which is given entirely by the second
term of eq. (\ref{QEDF}) above. Taking the infinite volume continuum
limit so that we may replace the sum over $\vec k$ by an integral, and
dividing by the volume, we find the usual result for the free energy
density of the photon gas, {\it viz.}
\begin{equation}
2 T \int {d^3 \vec k\over (2\pi)^3} \ln \bigl( 1 - e^{-\omega_k
\beta}\bigr) = - {\pi^2 \over 45} T^4 \ ,
\end{equation}
from which all the other usual thermodynamic relations follow. If we
had considered Yang-Mills theory with gauge group $SU(N)$ then this
same result holds to lowest order in the coupling $g$ if multiplied by
the number of independent fields in the adjoint representation of
$SU(N)$, namely $N^2 -1$. This is the leading contribution to the free
energy density of gluons in QCD (where $N=3$) at high temperature.

Thus, the covariant, geometrically constructed path integral of the
last section, together with a covariant definition of the functional
determinants by means of the $\zeta$-function, yields the same result
for the finite temperature free photon or gluon gas as that obtained
by the more standard (but non-covariant) canonical quantization in
Fock space. The Bose-Einstein statistics of the logarithm in
(\ref{QEDF}) comes out of the covariant treatment automatically and
does not require any additional assumptions. One can treat fermionic
degrees of freedom analogously by the introduction of anti-commuting
Grassman variables into the path integral, and obtain Fermi-Dirac
statistics for the corresponding thermodynamic functions of a gas of
spin-$1\over 2$ particles. These results may be obtained very rapidly within
the covariant path integral framework.

\section{The Functional Integral Over Geometries}

In the path integral approach to quantizing gravity a very important
technical issue that must be faced is the correct specification of the
functional measure on the coset space of spacetime manifolds
modulo coordinate reparameterizations. Although this problem has been
solved, originally by following the algebraic method based on the
Becchi-Rouet-Stora-Tyutin (BRST) global supersymmetry of the gauge
fixed path integral \cite{Fuj}, the issue of the functional measure in the path
integral for quantum gravity has continued to be a source of confusion.
Accordingly, it is worthwhile to rederive the BRST result by the more intuitive
geometric method of constructing the invariant volume element on the cotangent
space of metrics. The two methods lead to the same result \cite{BBM}.

One begins by recognizing that the action of classical general
relativity (or more generally of any theory consistent with the
principle of equivalence) is invariant under general coordinate
transformations, $x^{\mu} \rightarrow X^{\mu} = X^{\mu} (x) $ of the
manifold of spacetime. The infinitesimal form of this transformation
is
\begin{equation}
x^{\mu} \rightarrow x^{\mu} + \xi^{\mu} (x).
\label{coordtrans}
\end{equation}
We shall construct the functional measure in the path integral for
quantum gravity by requiring that it also be invariant under
(\ref{coordtrans}), following the same strategy as in the previous
examples \cite{BBM}. Treat an arbitrary spacetime metric $g_{\mu
\nu}(x)$ as the coordinate of a ``point" in the function space of all
metrics, denoted by $\cal M$. The infinitesimal one-form, $\delta
g_{\mu \nu}(x) \equiv h_{\mu \nu}(x)$ lies in the cotangent space to
$\cal M$ at the point $g_{\mu \nu}(x)$. We define a quadratic inner
product on this cotangent space of one-forms by
\begin{equation}
\langle h,h\rangle _{_T} \equiv \int d^4 x \sqrt{-g}\  h_{\mu \nu}(x)G^{\mu \nu
\rho \sigma}
h_{\rho \sigma}(x). \label{quadform}
\end{equation}
The subscript $T$ reminds us that this is an inner product for
tensors.  Now the scalar $ds^2 = g_{\mu \nu}(x)dx^{\mu} dx^{\nu}$ is
invariant under the (passive) relabelling of the coordinates of
spacetime, (\ref{coordtrans}) that leaves the geometric spacetime
point unchanged. The corresponding transformation of the metric on the
spacetime manifold is
\begin{equation}
h_{\mu \nu}(x) \rightarrow h_{\mu \nu}(x) + \nabla_{\mu} \xi_{\nu} +
\nabla_{\nu} \xi_{\mu}. \label{metrtrans}
\end{equation}
This may now be regarded as a relabelling of coordinates on $\cal M$
which leaves the point, {\it i.e.} the geometry corresponding to
$g_{\mu \nu}(x)$ unchanged. Hence we must require that the measure be
invariant under the transformations (\ref{coordtrans}) and
(\ref{metrtrans}) as well. Since $h_{\mu \nu}(x)$ transforms
covariantly as a symmetric tensor under (\ref{coordtrans}), $G^{\mu
\nu \rho \sigma}$ must transform like a contravariant four-tensor.
Like $g_{\mu \nu}(x)$, $G^{\mu \nu \rho \sigma}(g)$ has evident
symmetry properties: it is symmetric under interchange of its first
two indices or its last two indices, as well as interchange of the
first two with the last two.  Finally, again like $g_{\mu \nu}(x)$,
$G^{\mu \nu \rho \sigma}(g)$ must be a purely local function of the
coordinates of $\cal M$. That is, it should contain no derivatives of
$g_{\mu \nu}(x)$. The unique ultralocal tensors with these properties
are \cite{DeW}
\begin{equation}{ {1 \over 2}} (g^{\mu \rho} g^{\nu \sigma} + g^{\mu \sigma}
g^{\nu \rho})\qquad {\rm and}\qquad g^{\mu \nu} g^{\rho \sigma} \ .
\end{equation}

If we did not demand ultralocality, an infinite number of tensors
involving higher derivatives would appear in this list. Using such
tensors in the definition of the inner product and functional measure
ultimately would have the effect of defining a different set of
dynamical coordinates for the theory.  Since we assume that the metric
is the fundamental field coordinate, and derivatives of it in the
action introduce dynamics, we do not wish to introduce derivatives and
spurious dynamics into the essentially {\it kinematic} definition of
the inner product or functional measure. In fact, this is the only
principle which justifies an otherwise quite arbitrary distinguishing
of the functional measure in the path integral from the action
functional.

Restricting the metric on $\cal M$ to be covariant and ultralocal
determines it (up to an overall irrelevant normalization) to be:
\begin{equation}
G^{\mu \nu \rho \sigma} = { {1 \over 2}} (g^{\mu \rho}
g^{\nu \sigma} + g^{\mu \sigma} g^{\nu \rho} + C g^{\mu \nu} g^{\rho
\sigma}), \label{metr}
\end{equation}
where $C$ is an undetermined constant.

Having endowed the function space of metrics $\cal M$ itself with a
metric, (the ``supermetric" in the language of De Witt \cite{DeW}) we are now
in
a position to define an invariant volume form on the space. In order
to avoid confusion it is useful to introduce the vielbein field,
$e_{\mu}^m$ which converts spacetime vector indices to spacetime
tangent space indices. If we then define the density,
\begin{equation}
\tilde h_{mn} \equiv \sqrt{e}\, e_m^{\mu}\, e_n^{\nu}\, h_{\mu \nu},
\end{equation}
together with the relations,
\begin{equation}
g_{\mu \nu}= e_{\mu}^m\, e_{\nu}^n\, \eta_{mn},
\end{equation}
and
\begin{equation}
e\equiv \det (e_{\mu}^m) = \sqrt {-g},
\end{equation}
then the inner product (\ref{quadform}) may be expressed in terms of
the {\it flat} spacetime metric, $\eta_{mn} = {\rm diag}(-1, 1, 1, 1)$
as
\begin{equation}
\langle h,h\rangle _{_T} = \int d^4x \ \tilde h_{mn} \tilde G^{mnrs} \tilde
h_{rs},
\end{equation}
where
\begin{equation}
\tilde G ^{mnrs} = {\scriptstyle{\frac{1}{2}}}(\eta^{mr} \eta ^{ns} + \eta
^{ms} \eta ^{nr} +
C \eta ^{mn} \eta ^{rs})
\end{equation}
is independent of $x$.

Then by analogy with the invariant volume form on a pseudo-Riemannian
manifold, $\prod_m d\tilde x^m = \det (e_{\mu}^m) \prod_{\mu}
dx^{\mu}$, we would like to define the invariant volume form on the
function space of metrics $\cal M$ by:
\begin{equation}
[{\cal D} h_{\mu \nu}] \equiv const. \times \prod_{x,m \leq n} d\tilde
h_{mn}(x) = const. \times \prod_x e^{{(D-4)(D+1) \over 4}}\prod_{\mu
\leq \nu} dh_{\mu \nu}(x)
\label{measa}
\end{equation}
in $D$ spacetime dimensions. Although useful for some formal
manipulations this ``definition" leads to ambiguities for continuum
functional integrals, because of the ill-defined nature of the product
at each spacetime point, which also leaves the normalization of the
measure undefined. One would again have to resort to some kind of
discretization and limiting procedure in order to give meaning to
(\ref{measa}). More than just a technical inconvenience such a
procedure would be in grave danger of violating general covariance by
imposing a rigid skeletonization on a dynamical theory of spacetime,
which should have no {\it a priori} preferred geometry inherent in its
definition. Instead, as we have seen in the previous
examples, it is preferable to define the measure by the Gaussian
normalization condition,
\begin{equation}
\int [{\cal D}h_{\mu \nu}] \exp \bigl(-{\textstyle\frac{i}{2}}\langle
h,h\rangle _{_T} \bigr) = 1\ ,
\label{measb}
\end{equation}
which is formally satisfied by (\ref{measa}).

The overall constant in front of the supermetric, (\ref{metr}) is
irrelevant, since it may always be reabsorbed into the normalization
integral, (\ref{measb}).  The constant $C$ is {\it not} irrelevant,
since it determines the signature of the supermetric on $\cal M$. This
may be seen by decomposing the arbitrary tangent space tensor into its
tracefree and trace parts,
\begin{equation}
h_{\mu \nu} = h^{TF}_{\mu \nu} + {h\, g_{\mu \nu} \over 4}\ ,
\end{equation}
and operating on this decomposition with the supermetric $G$. The
$(D+2)(D-1) \over 2$ tracefree parts of $h_{\mu \nu}$ in $D$
dimensions are mapped onto $h^{TF \mu \nu}$, independent of $C$.
However, on the scalar trace mode $G$ has eigenvalue, $1 + {CD \over
2}$. Hence the signature of $G$ depends on the value of $C$: for $C >
-{D \over 2}$, the signature of $G$ in the scalar trace sector is
positive, while for $C < -{D \over 2}$ the signature is negative. If
$C = -{D \over 2}$, the metric is non-invertible and becomes a
projector onto the tracefree subspace. For the moment let us leave $C$
undetermined.

For any value of $C$ it is clear that the functional measure defined
with reference to the inner product, $\langle h,h\rangle _{_T}$ is
invariant under the infinitesimal general coordinate transformation,
(\ref{coordtrans}). Once we have a coordinate invariant functional
measure, we must extract the infinite gauge orbit volume in an
invariant way as well. To this end, we introduce a change of
coordinates in the tangent space of $\cal M$ at $g_{\mu \nu}$:
\begin{equation}
h_{\mu \nu} = h^{\perp}_{\mu \nu} + (L \xi)_{\mu \nu} + (2 \sigma +
{\textstyle {2 \over D}}\nabla_{\lambda} \xi^{\lambda})g_{\mu \nu},
\label{decomp}
\end{equation}
where $L$, (the ``conformal Killing form") maps vectors into traceless
symmetric tensors,
\begin{equation}
(L \xi)_{\mu \nu} \equiv \nabla_{\mu} \xi_{\nu} + \nabla_{\mu}
\xi_{\nu} - {\textstyle {2 \over D}} (\nabla_{\lambda}
\xi^{\lambda}) g_{\mu \nu}.
\label{Ldef}
\end{equation}
Thus, $L \xi$ spans all symmetric tensors which are gauge transforms
of $h^{TF}_{\mu \nu}$, the traceless part of $h_{\mu \nu}$.  The
scalar $\sigma$ is the gauge invariant piece of the trace, and
$h^{\perp}_{\mu \nu}$ is the gauge invariant piece of $h^{TF}_{\mu
\nu}$. Now $h^{\perp}_{\mu \nu}$ may be chosen to lie in the
orthogonal complement to $L$, with respect to the inner product
(\ref{quadform}), which requires $(L^{\dag} h^{\perp})_{\mu} =
-2\nabla^{\nu} h^{\perp}_{\mu \nu} = 0$.  Indeed this is the simplest
choice for doing one-loop calculations, \cite{BBM} which involve only
Gaussian integrals, and justifies the notation $h^{\perp}$.  However,
the choice of orthogonal coordinates on the tangent space of $\cal M$
is by no means necessary, and $h^{\perp}_{\mu \nu}$ may be required to
satisfy an arbitrary coordinate (gauge) condition:
\begin{equation}
(F\cdot h^{\perp})_{\mu} \equiv F^{\nu} h^{\perp}_{\mu \nu} = 0.
\label{cond}
\end{equation}
The only condition on $F$ is that the operator $F \circ L$ be locally
invertible, so that (\ref{decomp}) can be solved uniquely for $\xi$:
\begin{equation}
\xi_{\mu} = (F \circ L)^{-1 \ \nu} _{\mu}(F \circ h^{TF})_{\nu}.
\label{inver}
\end{equation}
Otherwise, the local coordinate chart, (\ref{decomp}) is singular at
the point $g_{\mu \nu}$.
\par
Following the discussion in the previous section for the case of the
non-abelian gauge field, to extract the infinite gauge orbit volume
generated by the gauge direction, $\xi_{\mu}$ we must find the
Jacobian of the transformation to the new field coordinates,
$(h^{\perp}_{\mu \nu}, \xi_{\mu}, \sigma)$:
\begin{equation}
[{\cal D}h_{\mu \nu}] = J\, [{\cal D}h^{\perp}_{\mu \nu}] [{\cal
D}\xi_{\mu}][{\cal D} \sigma].
\label{measfac}
\end{equation}
This is accomplished by substituting the decomposition, (\ref{decomp})
into the inner product (\ref{quadform}), completing the square of the
term quadratic in $\xi_{\mu}$, and computing the Gaussian integrals
over each of the components respectively:
\begin{eqnarray}
1 &=& \int [{\cal D}h_{\mu \nu}] \exp \bigl(- {\textstyle\frac{i}{2}} \langle
h,h\rangle _{_T}\bigr) \nonumber \\
&=& J \int [{\cal D}h^{\perp}_{\mu \nu}] \exp
\bigl( - {\textstyle\frac{i}{2}} \langle h^{\perp}, (1-M)h^{\perp}\rangle
_{_T}\bigr) \int [{\cal D}\xi_{\mu}] \exp \bigl( -
{\textstyle\frac{i}{2}}\langle \xi',\Delta_1 \xi'\rangle _{_V} \bigr)\nonumber
\\
& &\null \hskip 2 cm \times \int [{\cal D}\sigma] \exp \bigl( -8i(1+2C) \langle
\sigma,\sigma\rangle _{_S} \bigr),    \label{chan}
\end{eqnarray}
where the vector Laplacian $\Delta_1$ is defined by:
\begin{equation}
(\Delta_1)_{\mu}^{\ \nu} \equiv (L^{\dag} L)_{\mu}^{\ \nu} =
-2\Bigl(\delta_{\mu}^{\ \nu} \nabla^2 + ( 1 - {\textstyle\frac{2}{D}})
\nabla_{\mu} \nabla^{\nu} + R_{\mu}^{\ \nu} \Bigr)\ ,
\end{equation}
the tensor operator $M$ is given by:
\begin{equation}
M_{\mu \nu}^{\ \ \rho \sigma} \equiv \bigl(L (\Delta_1)^{-1}
L^{\dag}\bigr)_{\mu \nu}^ {\ \ \rho \sigma} = -2 \nabla_{\mu}
\bigl(\Delta_1^{-1}\bigr)^{\ \rho}_{\nu}\nabla^ {\sigma} -
2\nabla_{\nu} \bigl(\Delta_1^{-1}\bigr)^{\ \rho}_{\mu}\nabla^ {\sigma}
+ g_{\mu \nu}\nabla^{\lambda} \bigl(\Delta_1^{-1}\bigr)^{\ \rho}_
{\lambda}\nabla^{\sigma},
\end{equation}
and $\xi_{\mu}' = \xi_{\mu} + (\Delta_1^{-1}L^{\dag}h^{\perp})_{\mu}$
is the shifted vector obtained by completing the square.  The
notations $\langle ,\rangle _V$ and $\langle ,\rangle _S$ denote the
covariant inner products on vectors and scalars respectively:
\begin{equation}
\langle \xi,\xi\rangle _{_V} = \int d^4 x\, \sqrt{-g}\, \xi_{\mu} \,g^{\mu
\nu}\, \xi_{\nu},\label{vecinnr}
\end{equation}
\begin{equation}
\langle \sigma,\sigma\rangle _{_S} = \int d^4 x\, \sqrt{-g}\, \sigma^2
\end{equation}

The remaining tensor and vector Gaussian functional integrals in
(\ref{chan}) can be simplified \cite{BBM} to yield finally the result
for the Jacobian:
\begin{equation}
J = \bigl[\det{}_T (1 - M)\big\vert_{_F}\bigr]^{1 \over 2} \bigl[ \det{}_V
\Delta_1 \bigr]^{1 \over 2} = [\det{}_V (F\circ F^{\dag})]^{-{1 \over
2}}\det{}_V (F \circ L), \label{jacob}
\end{equation}
in complete analogy with the gauge theory result (\ref{YMjacobianb}).
The second factor will be recognized as the Fadeev-Popov determinant
for the gauge (\ref{cond}) on the tracefree components of $h_{\mu
\nu}$, while the first factor is a $h_{\mu \nu}$-independent
normalization factor that makes no contribution to the Feynman rules.
Notice that the Jacobian factor has been derived by tangent space
methods, involving only Gaussian functional integrals, but that this
involves no restriction to one-loop order. The result (\ref{jacob}) is
valid to {\it all} orders of perturbation theory.

With the Jacobian (\ref{jacob}), we now know how to factor the
infinite diffeomorphism gauge group volume out of the covariant
quantum measure, (\ref{measfac}) in a manifestly covariant way. If the
action is independent of the vector gauge orbit parameter $\xi_{\mu}$,
integration over $\xi_{\mu}$ would simply yield the infinite volume of
the diffeomorphism group, so
\begin{eqnarray}
\{Vol({\cal G})\}^{-1}\int [{\cal D}h_{\mu \nu}] &=& \{Vol({\cal G})\}^{-1}
\int [{\cal D}\xi_{\mu}] \int\, J\, [{\cal D}h^{\perp}_{\mu \nu}][{\cal
D}\sigma]
\nonumber \\ &=& \int\, J\, [{\cal D}h^{\perp}_{\mu \nu}][{\cal D}\sigma],
\end{eqnarray}
when integrated over functions independent of $\xi_{\mu}$.

The final steps in constructing the covariant path integral for
quantum gravity involves extending the integration measure defined on
the tangent space to a measure on the full metric. The extension of
the coordinates on the tangent space (\ref{decomp}) to coordinates on
$\cal M$ is straightforward, at least locally. We write
\begin{eqnarray}
g_{\mu \nu}(x) &=& {\partial X^{\rho} \over \partial x^{\mu}} {\partial
X^{\sigma} \over \partial x^{\nu}} e^{2 \sigma (X)} g_{\rho
\sigma}^{\perp}(X), \nonumber \\
(F\cdot g^{\perp})_{\mu} &=& 0.
\label{gperpdef}
\end{eqnarray}
where $\sigma$ may be fixed by the requirement that $g^{\perp}$ has
constant scalar curvature ($R^{\perp} = const.$):
\begin{equation}
R[g_{\mu \nu}] = 6e^{-3\sigma}\Delta_0^{\perp} e^{\sigma} +
e^{2\sigma} R^{\perp}
\label{yamabe}
\end{equation}
and $\Delta_0^{\perp}$ is the scalar Laplace-Beltrami operator with
respect to the metric $g^{\perp}$. The vacuum amplitude for quantum
gravity may then be written in the succint generally covariant form:
\begin{eqnarray}
Z &=& \{Vol({\cal G})\}^{-1} \int [{\cal D}g_{\mu \nu}] \exp \bigl( i
S_{inv}[g] \bigr)\nonumber \\
&=& \int\, J\,[{\cal D}g^{\perp}_{\mu \nu}] [{\cal D} \sigma] \exp \bigl( i
S_{inv} [ e^{2 \sigma} g^{\perp}]), \nonumber \\
&=&[\det{}_V (F\circ F^{\dag})]^{-{1 \over 2}}\ \int [{\cal D}\sigma] [{\cal D}
g^{\perp}_{\mu \nu}] \det{}_{_V} (F\circ L)\Big|_{g= e^{2\sigma}g^{\perp}}
\ \exp \bigl( i S_{inv} [ e^{2 \sigma}g^{\perp}]),
\label{Zgrav}
\end{eqnarray}
where (\ref{jacob}) has been used. This, together with Eqs. (\ref{Ldef}),
(\ref{gperpdef}), and (\ref{yamabe}) is the final result for the invariant
functional integral over geometries.

{\it Note that the infinite volume of the non-compact group of spacetime
coordinate transformations does not appear in} (\ref{Zgrav}), {\it and that the
functional measure in the invariant integration over} $\sigma$ {\it and}
$g^{\perp}$ {\it is determined completely by the Gaussian normalization
condition} (\ref{measb}) {\it on the cotangent space in the decomposition}
(\ref{decomp}).

If we set
\begin{equation}
g^{\perp}_{\mu \nu} = \bar g_{\mu \nu} + h^{\perp}_{\mu \nu},
\end{equation}
the analogy with the gauge theory result (\ref{YMfinal}) will be
evident. In both cases the $a^{\perp}$ or $h^{\perp}$ independent
determinant makes no contribution to the Feynman rules. It arises only
because
\begin{equation}
[{\cal D}h_{\mu \nu}] \delta (F \cdot h) = [\det{}_V (F\circ F^{\dag})]^{-{1
\over 2}} \ [{\cal D}h^{\perp}_{\mu \nu}],
\end{equation}
when both the $[{\cal D}h_{\mu \nu}]$ and $[{\cal D}h^{\perp}_{\mu \nu}]$
measures are normalized by Gaussian conditions (\ref{measb}) in their
respective spaces.

An interesting mathematical aspect of this construction of the
covariant measure in quantum gravity is the natural appearance of the
Yamabe condition (\ref{yamabe}) to fix completely the coordinates
$g^{\perp}$ and $\sigma$ in the manifold of physical metric fields
${\cal M}/{\cal G}$. This problem of classical Riemannian
geometry thus takes on an added importance at the quantum level. It
would be very interesting to know, for example, if there are global
obstructions to defining this coordinate system on the superspace
${\cal M}/ {\cal G}$, since such obstructions would contain
information about the global topology of the superspace which would
have implications for the quantum theory in a non-perturbative domain.
This is an area of mathematical investigation that has not been much
explored, and which remains open for future research.

With the above methods of introducing a supermetric and volume element
on the space of field configurations the path integral may be defined
for systems with gauge and/or reparameterization invariance. It cannot
be emphasized too strongly that in cases where the invariance group is
non-compact, naive discretization methods fail completely and it is
important to have some definition of the path integral where the
integration is over gauge equivalence classes of configurations only.
In the next section we shall consider the applications of this
geometric method of defining the continuum path integral in both two
and four dimensions.

\section{Applications of Covariant Functional Integration}

\subsection{The Conformal Anomaly in $2$ Dimensions}

A simple application of the covariant path integral is the calculation
of the trace anomaly in two dimensions, which leads to the
Polyakov-Liouville action for quantum gravity in two dimensions. For this
application we consider the invariant functional measure not over geometries
but over matter fields. For example, consider a single free massless scalar
field with the classical action,
\begin{equation}
S_{cl}[g, \phi] = \int d^2 x \, \sqrt {-g}\, g^{\mu\nu}
\partial_{\mu}\phi \partial_{\nu}\phi\ ,
\end{equation}
which is clearly invariant under general coordinate transformations.
In addition, it is also invariant under Weyl rescalings, since writing
\begin{equation}
g_{ab} = e^{2\sigma} \eta_{ab}\ ;\qquad \sqrt {-g} = e^{2\sigma}
\end{equation}
we observe that $S[\phi]$ is independent of $\sigma$. General
coordinate invariance implies that the energy momentum tensor derived
from $S[\phi]$ is covariantly conserved,
\begin{equation}
\nabla_aT^{ab}[g, \phi] = 0 \ ;\qquad T^{ab}[g, \phi] ={2 \over \sqrt {-g}}
{\delta S[g, \phi] \over \delta g_{ab}} = \partial_a\phi \partial_b\phi -
{\scriptstyle{\frac{1}{2}}} (\partial\phi)^2 g_{ab}
\end{equation}
while Weyl invariance guarantees that this classical energy momentum
tensor is traceless,
\begin{equation}
g_{\mu\nu}T^{\mu\nu}[\phi] = e^{-2\sigma}{\delta S\over \delta
\sigma}[g=e^{2\sigma}\eta, \phi] = 0\ .
\end{equation}

Now we define the quantum effective action (in Euclidean time) by the
covariant path integral,
\begin{equation}
Z = \exp (- S_{eff}[g]) = \int [{\cal D}\phi] \exp (-S_{cl}[g, \phi])
\label{sSeff}
\end{equation}
where the generally covariant integration measure over scalar fields
must be defined such that
\begin{equation}
\int [{\cal D}\phi] \exp (-{\textstyle\frac{1}{2}}<\phi, \phi>) =1\ ; \qquad
<\phi, \phi> \equiv \int d^2 x\, \sqrt {+g}\, \phi^2\ .
\label{smeas}
\end{equation}
Now the point is that this inner product and the corresponding
integration measure over scalar fields $[{\cal D}\phi]$ is invariant
under general coordinate transformations but {\it not} under Weyl
rescalings. Hence we must expect that the energy-momentum tensor of
the quantized theory will remain conserved but have non-zero trace.
Although this had to be discovered by laborious calculations in the
standard operator quantization method, and for this reason was called
the trace anomaly \cite{DesDufIsh}, it is actually obvious from the Weyl
non-invariance of the covariant integration measure. Notice that it is
logically possible to define a Weyl invariant scalar inner product and
integration measure by leaving out the $\sqrt {+g}$ in (\ref{smeas})
above, at the price of making it not generally coordinate invariant.
In that case the quantum energy momentum tensor would remain
traceless, but it would no longer be conserved. If coordinate
invariance is assumed to be a more fundamental symmetry of nature than
Weyl invariance this possiblity must be rejected.

In order to calculate the Weyl trace anomaly we perform the Gaussian
integration in (\ref{sSeff}) and obtain
\begin{equation}
S_{eff}[g] = {\textstyle\frac{1}{2}} {\rm tr} \ln (-\sq) \rightarrow
-{\textstyle\frac{1}{2}}\int_{\epsilon}^{\infty} {dt\over t}{\rm tr} \exp (- t
\sq)
\end{equation}
where we have used the heat kernel definition of tr ln and
introduced a cutoff on the lower limit of integration. This regulated
form is most convenient for evaluating the trace anomaly by varying
with respect to $\sigma$. Using
\begin{equation}
\sq = {1\over \sqrt g}\partial_{\mu}[\sqrt g g^{\mu \nu}\partial_{\nu}] =
e^{-2\sigma} \sqb
\label{sop}
\end{equation}
with $\sqb$ evaluated in the flat Euclidean metric, we find
\begin{eqnarray}
g_{ab}T^{ab}_{qu} &=& e^{-2\sigma}{\delta S_{eff}[g]\over
\delta\sigma(x)} \nonumber \\
&=& - \int_{\epsilon}^{\infty} dt\,\langle x\vert \sq e^{-t\sq}\vert x\rangle
\nonumber \\ &=& \langle x\vert e^{-\epsilon\sq}\vert x\rangle
\nonumber \\ &=& {1\over 4\pi}\Biggl[ {1\over \epsilon} + {R \over 6} + {\cal
O}(\epsilon)\Biggr]\ ,
\label{tranom}
\end{eqnarray}
where the expansion of the heat kernel for $e^{-\epsilon\sq}$ has been
used in the last step as $\epsilon \rightarrow 0$, and we have assumed
that the operator $\sq$ has no zero modes so that the upper limit of
the $t$ integral does not contribute. The background metric
independent, divergent first term is again associated with the
infinite energy density of the vacuum in flat space which can be
regulated by the full $\zeta$-function method, or simply subtracted
from the definition of the energy momentum tensor by a normal ordering
procedure. The finite second term proportional to the Ricci scalar
curvature of the background metric $g_{ab}$ is the trace anomaly.
Notice that it indeed comes from the Weyl noninvariance of the
integration measure (\ref{smeas}), since if we used the Weyl invariant
measure omitting $\sqrt g$ from the inner product defined there, $\sq$
in (\ref{sop}) above would be replaced by $\sqrt g \sq = e^{2\sigma} \sq
= \sqb$ which is independent of $\sigma$, so that the variation in
(\ref{tranom}) would then give zero identically. The trace ``anomaly"
therefore is a necessary and immediate consequence of the covariant
definition of the path integral measure in (\ref{smeas}).

\subsection{The Path Integral for Quantum Gravity in Two Dimensions}

In the previous subsection the trace anomaly for one ($N_S =1$) scalar
field was calculated by the heat kernel method. The general form of
the trace anomaly of the energy momentum tensor for classically Weyl
invariant matter in a background gravitational field is
\begin{eqnarray}
T_a^{a \ (matter)} &=& {c_m\over 24 \pi} R \\
&=& {c_m\over 24
\pi} e^{-2 \sigma} (\overline R - 2\sqb\sigma), \qquad D=2
\end{eqnarray}
in the decomposition $g_{ab} = e^{2\sigma} \bar g_{ab}$. The
coefficient $c_m = (N_S + N_F)$ for $N_S$ scalar and $N_F$ (Dirac)
fermion fields. From eq. (\ref{tranom}) this implies that there exists
an effective anomalous quantum action such that
\begin{equation}
{\delta S_{anom}[g]\over \delta\sigma (x)} = {c_m\over 24 \pi} (\overline R -
2\sqb\sigma)\ ,
\end{equation}
over and above any classical action for $\sigma$. Since the right side
of this equation is linear in $\sigma$ we may integrate both sides
immediately with respect to $\sigma$ to obtain the anomalous action:
\begin{equation}
S_{anom}[g=e^{2\sigma}\bar g] = S_{anom}[\bar g] + {c_m\over 24 \pi} \int d^2 x
\ \sqrt {{-\bar g}}\bigl[ - \sigma \sqb \sigma + \overline R \sigma \bigr] .
\label{PLloc}
\end{equation}
Since $S_{anom}[g]$ must be a scalar under general coordinate
transformations and a functional of only the full $g_{ab}$, we may
use this information to determine the $\sigma$ independent integration
constant $S_{anom}[\bar g]$ and write down the fully covariant but non-local
form of the anomalous action:
\begin{equation}
S_{anom} = -{c_m \over 96 \pi}\int d^2x \sqrt {-g}\int d^2x'\sqrt{-g'}
\ R(x)
\sq ^{-1}(x,x') \ R(x')\ .
\label{loctwo}
\end{equation}
If one adds to this induced action the classical Einstein-Hilbert
action in two dimensions,
\begin{eqnarray}
S_{cl} &=& \int d^2x \sqrt {-g} \ (\gamma R - 2\lambda)\nonumber \\
&=& 4 \pi \gamma \chi - 2 \lambda \int d^2x\ \sqrt {{-\bar g}} e^{2\sigma} ,
\end{eqnarray}
one obtains the Polyakov-Liouville action which describes the
fluctuations of random geometries, {\it i.e.} quantum gravity in two
spacetime dimensions.  Notice that the Einstein term alone describes no
dynamics, since the integral of the scalar curvature is just $\chi$,
the Euler number, and a topological invariant in $D=2$. Hence its
local variation vanishes and one would obtain no equation of motion at
all for the metric at the classical level. However, the addition of
the anomalous action $S_{anom}[g]$ changes things dramatically. In its
local form (\ref{PLloc}) we observe that a kinetic term for the $\sigma$
part of the metric has been generated by the path integration over the
matter fields, and the dynamics of the metric can now be non-trivial.

Eq. (\ref{Zgrav}) defines the covariant path integral for quantum
gravity in a general number of spacetime dimensions. Restricting to
$D=2$ Euclidean dimensions we observe that if the matter fields are considered
as
coordinates of a target space with dimension $c_m$, then we may also
view the same expression as the partition function for non-critical
closed bosonic string theory with a fixed world sheet topology.
Summing $Z$ over the number of handles $h$ yields the full generating
function for closed string amplitudes. It is the starting point for
all of the modern analysis of string amplitudes in terms of the
geometry of Riemann surfaces. The importance of the covariant point of
view in elucidating the properties of string theory in terms of
Riemannian geometry has been emphasized in the literature \cite{DhP}. In the
literature orthogonal coordinates on the space of metrics is generally
used which implies $F= L^{\dagger}$, although this is by no means
necessary. In these coordinates eq. (\ref{Zgrav}) becomes
\begin{equation}
Z = \int [{\cal D}\sigma] [{\cal D}g^{\perp}_{\mu \nu}] [\det{}_V (L\circ
L^{\dag})]^{+{1 \over 2}}\ \Big|_{g= e^{2\sigma}g^{\perp}}
\ \exp \bigl( - S_{inv} [ g=e^{2 \sigma}g^{\perp}])\ ,
\label{Zgravt}
\end{equation}
where for $S_{inv}$ we now put $-S_{anom} - S_{cl}$ above, after continuation
to Euclidean time.

In two dimensions the kernel of the operator $L^{\dagger}$ is finite
dimensional, being the space of quadratic differntials which is the
tangent space to the Teichm\"uller space $\cal T$. This is defined by
\begin{equation}
{\cal T} = {\cal M}_{const}/{\cal G}_0
\end{equation}
where ${\cal G}_0$ denotes the space of diffeomorphisms continuously
connected to the identity and
\begin{equation}
{\cal M}_{const} \equiv \{g^{\perp} ; R[g^{\perp}] \equiv \overline R = const
\}
\end{equation}
is the Yamabe slice corresponding to the metric $g$ obtained by solving
the uniformization (Yamabe) problem in two dimensions, namely \cite{Yam}
\begin{equation}
R[g= e^{2\sigma}g^{\perp}] = e^{-2\sigma}(\overline R - 2\sqb \sigma)\ .
\end{equation}
Since this always has a solution in $D=2$, the $\sigma$ and $g^{\perp}$
with constant $\overline R$ corresponding to any metric $g$ can always be
uniquely determined. In fact the constant $\overline R$ can always be chosen
to be $+1$ for two dimensional surfaces with $h=0$ handles (the
topology of $S^2$), $0$ for surfaces with $h=1$ handles (the topology
of the torus $T^2$), and $-1$ for surfaces of higher genus with $h \ge
2$. Then it is known that
\begin{equation}
{\rm dim ( ker} L^{\dag}) = {\rm dim}\, {\cal T}=
\cases {0, & $h=0$,\cr 2, & $h=1$,\cr 6h-6, & $h\ge 2$. \cr}
\end{equation}
The basis tensors of the finite dimensional Teichm\"uller space are
called moduli deformations, and may be denoted by $f_j$. Hence we may
expand
\begin{equation}
\delta g^{\perp}_{\mu\nu} = \sum_j \delta m_j (f_j)_{\mu\nu}
\end{equation}
where $\delta m_j$ denotes the modular deformation of the $j^{th}$
modular parameter over which we should finally integrate. Changing
integration variables from $\delta g^{\perp}_{\mu\nu}$ to $\delta m_j$
generates a finite dimensional determinant in the integration measure,
namely
\begin{equation}
[{\cal D}g^{\perp}_{\mu \nu}] = [{\rm det} \langle \phi_i \vert \phi_j
\rangle_{_T}]^{-{\frac{1}{2}}}{\rm det} \langle f_i
\vert\phi_j\rangle_{_T}\, \prod_j dm_j
\label{modul}
\end{equation}
where the $\phi_i$ are an arbitrary basis for ker $L^{\dag}$, and the
inner product for symmetric tensors is defined by Eq. (\ref{quadform})
for $D=2$. This is the analog of the Jacobian factors
(\ref{YMjacobianb}) or (\ref{jacob}) arising from the previous changes
of variables to an arbitrary basis in the quotient space ${\cal M}/
\cal G$. As in those cases it is possible (though not necessary) to
choose an orthogonal basis depending on the field point in $\cal M$ in
such a way as to reduce the Jacobian to a single determinant. Here the
choice $\phi_i = f_i$ accomplishes that. The advantage of keeping the
Jacobian in the more general form (\ref{modul}) is that the basis
$\phi_i$ may be chosen in such a way that $\langle f_i
\vert\phi_j\rangle$ is independent of the field point
$g_{\mu\nu}=e^{2\sigma}\bar g_{\mu\nu}$ and may removed from the
functional integral, so that all the field dependence in contained in
the first determinant alone.

If $h=0\,,1$ then there are also globally defined conformal Killing
fields obeying $L\psi = 0$, since
\begin{equation}
{\rm dim (ker} L) =
\cases {6, & $h=0$,\cr 2, & $h=1$,\cr 0, & $h\ge 2$ .\cr}
\end{equation}
in accordance with the Riemann-Roch index theorem,
\begin{equation}
{\rm dim (ker} L) - {\rm dim (ker} L^{\dag}) = 3\chi = 6 (1-h)\ .
\end{equation}
The existence of conformal Killing fields implies that the coordinate
chart $(\sigma, \xi , g^{\perp})$ on the space ${\cal M}$ is not
completely well-defined, since the decomposition (\ref{decomp}) can no
longer be inverted as in (\ref{inver}). Because this case is of some
interest let us consider it in more detail.

First we should distinguish conformal Killing vectors with zero
divergence, namely the Killing symmetries ($\nabla \cdot K_i = 0$,
KV's), from those with non-zero divergence, which we term proper
conformal Killing vectors ($\nabla \cdot \psi_i \ne 0$, PCKV's). The
latter generate diffeomorphisms which are equivalent to a shift in
$\sigma$. Hence the existence of PCKV's means that $\sigma$ has components
along the gauge fiber $\cal G$, and we have not fully realized our aim
of separating the gauge invariant field coordinates in ${\cal M}/ \cal
G$ from those along the fiber $\cal G$, or in physicist's language, we
have not completely fixed the gauge. This arbitariness may be remedied
easily enough by requiring that $\sigma$ be orthogonal to $\nabla \cdot
\psi_i$ for each PCKV. Then by a change of coordinates in this finite
dimensional space spanned by the $\psi_i$,
\begin{equation}
[{\cal D}\sigma]_{\psi} = \tilde J [{\cal D}\psi]\ ,
\label{PCKVJ}
\end{equation}
we find \cite{Vas}
\begin{equation}
\tilde J = [\det{}_V \langle \psi_i \vert \psi_j\rangle ]^{+{1 \over 2}}\ ,
\label{tildeJ}
\end{equation}
by a calculation exactly similar to the previous ones leading to
(\ref{YMjacobianb}) and (\ref{jacob}). In this way we can extract the
volume of the gauge fiber projected onto the PCKV's $[{\cal D}\psi]$
from the scalar part of the measure $[{\cal D}\sigma]$ in
(\ref{Zgravt}), and are left with integrations only over those $\sigma$
orthogonal to the fiber and the finite dimensional Jacobian $\tilde
J$.

The true Killing symmetries (unlike the PCKV's) generate {\it no}
change in the metric whatsover, so they cannot be treated in the same
manner. Indeed the existence of KV's implies not that there is some
residual gauge freedom in our coordinates $g^{\perp}$ and $\sigma$, but
that the dimension of the gauge fiber $\cal G$ has {\it decreased} by
precisely the number of linearly independent solutions of Killing's
equation,
\begin{equation}
\nabla_{\mu}K_{\nu} + \nabla_{\nu}K_{\mu} = 0\ .
\end{equation}
This is clearly a property of the geometry about which we are
expanding, and {\it not} the choice of coordinates of that expansion.
So, the KV's really should be treated differently from the PCKV's. In
the string theory literature, the practice has been to treat all the
CKV's on an equal footing, formally dividing $Z_2$ by the {\it
infinite} volume of the non-compact Lie group corresponding to the
full set of conformal Killing vectors ($SO(3,1) \simeq SL(2,C)$ in the
case of $S^2$), so that the vacuum amplitude actually {\it vanishes}
identically. Then, one shifts attention to {\it non-vacuum} matrix
elements of reparameterization invariant vertex operators which lead
to a compensating infinite factor in the numerator and finite results.
Although the results so obtained are evidently correct (since they
agree with older non-covariant operator calculations), this procedure
of dividing one infinity by another clearly is not completely
satisfactory, and it also leaves unanswered the question of how to
treat the true Killing symmetries correctly. Let us for the moment
ignore this subtlety and revisit the issue in more detail in the next
section on semi-classical methods.

In order to make use of the expression (\ref{Zgravt}) with
(\ref{modul}) the $\sigma$ dependence of all the factors must be
determined. This can again be done by the heat kernel method, the
details of which may be found in the literature \cite{DhP}. The result in $D=2$
is:
\begin{equation}
Z_2 = \int \prod_j dm_j \ J_2[{\cal D}\sigma]_{\bar g} \ e^{- cS_{anom} -
S_{cl}}\ ,
\label{string}
\end{equation}
where the total Jacobian is
\begin{equation}
J_2 = [{\rm det} \langle \bar \phi_i \vert \bar \phi_j
\rangle_{_T}]^{-{\frac{1}{2}}}\, {\rm det} \langle \bar f_i \vert \bar
\phi_j\rangle_{_T} \langle \bar \psi_i \vert \bar \psi_j
\rangle_{_V}]^{\frac{1}{2}} [\det{}'_V (\bar L^{\dag}\circ \bar L)]^{+{1
\over 2}}
\end{equation}
and
\begin{equation}
c = c_m - 26 + 1 = N_S + N_F - 25\ .
\label{central}
\end{equation}
The $-26$ comes from the heat kernel expansion of $L^{\dagger}L$ and
the additional $+1$ arises from the Weyl noninvariance of the $\sigma$
field itself which contributes to the anomaly coefficient just like
one additional scalar field. Thus, the net effect of the $\sigma$
dependence of the measure for matter plus metric in $D=2$ dimensions
is that we should replace $c_m$ by $c$ above, or in other words that
the gravitational contribution to the trace anomaly is of the same
form as that of the matter and would induce the Polyakov-Liouville
action even in the complete absence of matter fields. Again, this is a
purely quantum effect, arising in the path integral from the Weyl
noninvariance of the proper covariant integration measure.

\subsection{The Semi-classical Partition Function on de Sitter Space}

Let us begin this subsection with Lorentzian signature and the usual
Einstein-Hilbert action of general relativity,
\begin{equation}
S_{cl} = {1\over 16\pi G}\int _{\cal M}{d^4}x\sqrt{-g}\,(R - 2
\Lambda)
\end{equation}
augmented by a positive cosmological term. The Euler-Lagrange
variational principle $\delta S_{cl} = 0$ yields the equations of
motion,
\begin{eqnarray}
R_{ab}-{\textstyle\frac{1}{2}}R g_{ab}  &=& - {\Lambda}g_{ab} \nonumber \\
{\rm or} \qquad \qquad R_{ab} &=& {\Lambda}g_{ab}\ ,
\end{eqnarray}
which admits a maximally symmetric solution, de Sitter spacetime with
$O(4,1)$ isometry group. This classical solution is of particular
interest in early universe cosmology.

Consider now expanding the action $S_{cl}$ to second order in metric
fluctuations around the maximally symmetric de Sitter solution,
utilizing the decomposition (\ref{decomp}) and (\ref{Ldef}). We find
\begin{equation}
32{\pi}G\ {\delta}^2S = -<h^{\perp},{\Delta}_2 h^{\perp}>_{_T} - 24
<{\sigma},(-{\Delta}_0 +{\textstyle{1\over 3}}R){\sigma}>_{_S}\ ,
\label{soact}
\end{equation}
where ${\Delta}_2$ is the tensor Lichnerowicz Laplacian,
\begin{equation}
({\Delta}_2h)_{ab}={\Delta}_0h_{ab} +
2{{R_{a}}^{cd}}_{b}h_{cd} + {\scriptstyle{\frac{1}{2}}}Rh_{ab}\ ,
\end{equation}
and ${\Delta}_0$ is the ordinary scalar Laplacian, ${\Delta}_0=-{\nabla}^2$.
As expected, because of the general coordinate invariance of the
action, ${\delta}^2S$ does not depend on the deformations generated by
the diffeomorphisms ${\xi}$. However, the change of variables leaves
behind a relic, namely the Jacobian factor $J$ given by (\ref{jacob})
which becomes
\begin{equation}
J = \Bigl[\det{}_{_V}' (\bar L^{\dag}\circ\bar L)\Bigr]^{\frac{1}{2}}
\end{equation}
if orthogonal coordinates with respect to the inner product
(\ref{quadform}) are used to define $h^{\perp}$ and the CKV zero modes
are excluded from the $\det_{_V}'$. Making the further change of
variables in the $D+1=5$ dimensional space of PCKV's as described by
(\ref{PCKVJ}) and (\ref{tildeJ}) of the last subsection we obtain the
semiclassical Gaussian functional integral of small fluctuations in
Einstein gravity around de Sitter spacetime:
\begin{equation}
Z^{sc}[deS] =e^{iS_{cl}}\int J\tilde J[{\cal D}{\sigma}]'
\int [{\cal D}h^{\perp}]\,
{\rm exp}{\left({i\over 64{\pi}G}\left[-<h,{\Delta}_{2}\,h> + 24
<{\sigma}, (\Delta_0 - {\textstyle\frac{1}{3}}R){\sigma}>\right]\right)}\ .
\label{gaussdeS}
\end{equation}
Naively one would think that the neglect of the higher orders of
fluctuations in the exponent is justified in the limit of very weak
Newtonian coupling $G\Lambda\rightarrow 0$. We shall see in a moment
that this expectation will turn out to be {\it incorrect}.

Let us evaluate the Jacobian $J$ a bit more explicitly.  The
operator $\Delta_1 \equiv L^{\dag} L$ maps vectors into vectors.  The
space of all vectors may be decomposed into transverse and
longitudinal vectors according to
\begin{equation}
\xi_a = \xi^{\perp}_a + \nabla_a \phi
\end{equation}
A simple exercise in commuting covariant derivatives leads to the
results,
\begin{equation}
(\Delta_1 \xi^{\perp})_a = 2(\Delta_0 -{\textstyle\frac{1}{4}}R)\xi^{\perp}_a,
\end{equation}
and
\begin{equation}
(\Delta_1 \nabla \phi)_a = 3 \nabla_a (\Delta_0 - {\textstyle\frac{1}{3}}R)
\phi
\end{equation}
in the maximally symmetric de Sitter background.
This means that the Jacobian expressed as a vector determinant may be
rewritten as a product of determinants over transverse vectors
$\xi^{\perp}$ and scalars $\phi$:
\begin{equation}
J = \left[\det{}_{_V}' \bar\Delta_1\right]^{1 \over 2} = \left[\det{}_{_{\perp
V}}'(\Delta_0-{\textstyle\frac{1}{4}}R)\right] ^{1 \over 2}
\left[\det{}_{_S}''(\Delta_0 - {\textstyle \frac{1}{3}}R)\right]^{1 \over 2}
\end{equation}
where an irrelevant real multiplicative constant has been discarded.
The notation $\det''_S$ has been used to remind us that the original
determinant of $L^{\dag}L$ contains {\it no} zero modes or negative
modes, so such modes cannot be present in the scalar determinant on
the right side of the relation either. Notice otherwise that the
scalar operator appearing in this determinant, $\Delta_0 - {R \over
3}$ is the same as that appearing in the quadratic action (\ref{soact}).
Hence the Gaussian integration over $[{\cal D}{\sigma}]'$ in (\ref{gaussdeS})
would cancel completely against this part of the Jacobian in the
covariant measure, except for two subtleties. First, the {\it sign} of
the operator $\Delta_0 - \frac{R}{3}$ in (\ref{soact}) appears to be
different from that in the Jacobian, and second, the integration
$[{\cal D}{\sigma}]'$ excludes zero modes but {\it includes} possible negative
modes of the differential operator not included in the Jacobian, so
any cancellation will not be complete.

The meaning of the cancellation is of course the same as found in the
one loop partition function of QED, namely the role of the Jacobian is
to cancel unphysical modes in the covariant decomposition of
small fluctuations. In QED these are longitudinal modes which are
eliminated in a canonical framework by Gauss' Law constraint. In
Einstein gravity, one would expect the scalar modes of $\sigma$ to be
eliminated by the constraints of diffeomorphism invariance in a
canonical space plus time splitting. In the covariant functional
integral this elimination shows up in the cancellation of scalar
determinant by the correct Jacobian in the measure.

The question of the sign of the operator $\Delta_0 - \frac{1}{3}R$ in
(\ref{soact}) is related to the ``conformal factor problem" of Einstein
gravity which has received some attention (and caused a fair amount of
confusion) in the literature, particularly in the context of
continuation to Euclidean signature metrics \cite{Haw}. Then the ``wrong" sign
would mean the Gaussian integral becomes exponentially {\it divergent}
and consequently makes no sense at all. However, this is clearly an
artificial problem and not a real difficulty since in the canonical
approach to quantum gravity these scalar modes really are eliminated
by the constraints (at least in perturbation theory around flat space)
and cannot possibly lead to any physical instability in a correct
covariant treatment. The resolution proposed \cite{BBM} to this ``conformal
factor problem" is quite simple. In the definition of the inner product on the
cotangent space of metric deformations (\ref{quadform}) we required a metric
(\ref{metr}) which contained an undetermined constant $C$. Since $C$ is not
fixed by considerations of coordinate invariance alone, it can be determined
only by additional information about the particular invariant action $S_{inv}$
under consideration, {\it i.e.} it depends on dynamics and not pure kinematics,
and may vary from theory to theory. In the Einstein theory the kinetic term for
the $\sigma$ part of the action has the ``wrong" sign, but canonical
quantization methods precisely then imply that the constant $C =
-{ \frac{2}{3}} < -{ {\frac{1}{2}}}$ in De Witt's supermetric
(\ref{metr})\cite{DeW}. This means that the inner product
\begin{equation}
<{\delta}g,{\delta}g>\Big\vert_g= <h^{\perp}, h^{\perp}> + <\xi, L^{\dag}L
\xi> + {(1 + 2C) \over 4}<h, h>
\end{equation}
has a {\it negative} sign in its last term (in the scalar subsector of
trace modes). Hence the conformal factor scalar Gaussian integration
\begin{equation}
\int [{\cal D} \sigma] {\rm exp}\Biggl(-{i \over 2}{(1 + 2C) \over 4} <\sigma,
\sigma>\Biggr) = 1,
\end{equation}
is normalized with the {\it opposite} sign from the tensor modes
\begin{equation}
\int [{\cal D} h^{\perp}] {\rm exp} (- {\textstyle\frac{i}{2}} <h^{\perp},
h^{\perp}>)=1\ ,
\end{equation}
or in other words, the Euclidean continuation of the scalar modes
should be performed in the opposite sense from the tensor modes in the
semi-classical Einstein theory. This is the formal justification for the
otherwise quite {\it ad hoc} prescription proposed by Hawking to integrate over
imaginary conformal deformations \cite{Haw}. The covariant functional integral
makes
it clear that some choice of $C$ is necessary and why the naive continuation
to Euclidean signature is incorrect if $C < -{ {\frac{1}{2}}}$. A fully
consistent
derivation of this condition from the BFV path integral in phase space has yet
to be given however.

With the value of $C < -{ {\frac{1}{2}}}$ the Euclidean
continuation of the Gaussian integrations in (\ref{gaussdeS}) is
unambiguous and well-defined, and the scalar determinants in
(\ref{soact}) and $J$ do indeed cancel except for the subtlety of zero and
negative modes. The Euclidean continuation of de Sitter spacetime is just
$S^4$ with radius $a$ equal to $\sqrt {12\over R}$. Hence the scalar
operator $\Delta_0 - \frac{R}{3}$ has the discrete spectrum, $[n(n+3)
-4]/a^2$ with degeneracy $(2n +3)(n+1)(n+2)/6$. The five zero modes at
$n=1$ correspond precisely to the five PCKV's, $\nabla \cdot \psi_i \neq 0$,
$i=1,\dots 5$ which have been taken into account by the additional finite
dimensional Jacobian $\tilde J$ \cite{Vas}. The contribution of all of the
positive eigenvalue modes at $n= 2, 3, \dots$ are cancelled precisely
by the same modes in the Jacobian $J$. We are then left with only the
one negative eigenmode at $n=0$, corresponding to homogeneous
expanding or shrinking of the radius of $S^4$. This mode is not
eliminated by anything, and yields a {\it divergent} Gaussian
integration corresponding to a genuine instability of the de Sitter
background with respect to fluctuations in the average scalar
curvature. Notice that this instability is {\it infrared} in
character, having only to do with large scale fluctuations of the
metric, and nothing at all to do with ultrashort Planck scale physics.
For this reason, the results of the semiclassical analysis can be
trusted, notwithstanding the ultraviolet nonrenormalizability of
Einstein gravity. This infrared instability has been expected on other
grounds \cite{EM}, and is nicely confirmed in the covariant
functional integral approach.

Ignored in this discussion of (\ref{gaussdeS}) is the question raised
in the last subsection (in $D=2$) of how to treat the Killing vectors
of de Sitter spacetime, or its Euclidean continuation to $S^4$.
Non-trivial solutions to Killing's equation imply that a subgroup of
the gauge group of diffeomorphism invariance $\cal G$ acts trivially
at the field point possessing those Killing symmetries. Geometrically
this is easy to visualize if we think of a simple analogy with a
finite dimensional system possessing spherical symmetry. The symmetry
group is $O(N)$ in $N$ dimensions and generates a gauge orbit with
non-zero volume at every point of configuration space, except the
origin. Near the origin the spherically symmetric invariant volume
element goes to zero as $r^{N-1}dr$. At the origin it is identically
zero because the invariance group acts trivially there. In the
invariant path integral for quantum gravity, the factorization of the
gauge orbit volume from the measure around a generic metric produces
the Jacobian, det$(L^{\dag}L)$ which goes to zero when there are
Killing symmetries. Hence, properly speaking, the invariant path
integral measure around a metric with Killing symmetries actually {\it
vanishes} identically. This means that a naive semiclassical
evaluation of the functional integral neglecting the zero in the
measure at the symmetric point cannot be correct.

To show this let us pursue a bit further the simple analogy with
rotional invariance in a finite dimensional integral. Consider
evaluating
\begin{equation}
I(\epsilon) = 2\int_0^{\infty} \, dr\, r^{N-1} \exp \Bigl(-{r^2\over
\epsilon}\Bigr)\ ,
\end{equation}
where $\epsilon$ is assumed to be a small parameter. A naive steepest
descent evaluation of this integral would consider only the saddle
point at $r=0$, ignore the Jacobian measure factor and estimate $I
\simeq \epsilon^{\frac{1}{2}}$. This is obviously incorrect since the
integral is easily evaluated exactly:
\begin{equation}
I(\epsilon) = \epsilon^\frac{N}{2}\, \Gamma({\textstyle\frac{N}{2}})\ .
\end{equation}
A much more accurate saddle point estimate for the integral is
obtained by promoting the measure factor into the exponent and
treating $1/N$ and $\epsilon$ as small parameters of the same order.
Then the modified saddle point occurs at non-zero $r^2_0 =
(N-1)\epsilon /2$ and we obtain the estimate,
\begin{equation}
I \approx 2\, \epsilon^{\frac{N}{2}}\,
\bigl({\textstyle\frac{\pi}{2}}\bigr)^{\frac{1}{2}}\,
\bigl({\textstyle\frac{N-1}{2}}\bigr)^{\frac{N-1}{2}}\, e^{-\frac{N-1}{2}}\ ,
\end{equation}
which is clearly more accurate than the naive estimate, gives the
correct $\epsilon$ dependence for any $N$, and a numerical prefactor
which agrees with the exact answer in the limit $N\rightarrow \infty$.
Even for $N=2$ the saddle point estimate gives the numerical prefactor
$\sqrt {\pi\over e} = 1.075$ compared to the exact $\Gamma (1) = 1$.

The important lesson of this simple example is that it is not correct
to do the naive saddle point expansion around a point where the
measure vanishes, since the measure shifts the correct saddle point
away from the most symmetric point. Expanding the invariant path
integral for quantum gravity around a geometry with normalizable Killing
vectors
and ignoring the zero in the Jacobian measure is just such a mistake,
since the quantum measure vanishes there and must be taken into
account to find the correct {\it non-symmetric} saddle point. The de Sitter
geometry gives precisely {\it zero} contribution to the partition function for
quantum gravity because it has zero weight in the functional integral, when the
correct invariant measure is taken into account. This phenomenon is related to
the linearization instability of the classical solution in the canonical
approach to
classical gravity \cite{Mon}, whose interpretation in the full quantum theory
has remained somewhat obscure. Notice that there is no such problem in flat
space since the Killing vectors in that case are {\it not} normalizable with
respect to the inner product (\ref{vecinnr}) which defines the integration
measure on the fiber of gauge equivalence $\cal G$. Hence there is no
linearization or other
instability of flat spacetime \cite{MazMot}.

Finally, as we have seen, the $\sigma$ zero mode
did not cancel against anything and has a Gaussian with the sign corresponding
to instability of the de Sitter background. Hence the quantum state of
the gravitational field of linearized fluctuations around de Sitter space
cannot be de Sitter invariant. The consequences of this infrared behavior
in de Sitter space for quantum gravity in the large is the subject
of the next section. Once again we see that the invariant
functional integration measure has physical consequences which may be
exposed directly, without the cumbersome calculations of noncovariant
approaches.

\subsection{The Effective Theory of the Conformal Factor in Four Dimensions}

We have seen that the diffeomorphism constraints encoded in the Jacobian
of change of variables $J$ does not completely cancel the excitations of
$\sigma$
around de Sitter space, so that the quantum theory does not preserve
exactly the same diffeomorphism constraints as the classical Einstein
theory. The surviving mode is an infrared mode corresponding to dilation
of the entire space. The existence of an infrared instability in the quantum
fluctuations around de Sitter spacetime indicates that there is nontrivial
dynamics
in the conformal sector of four dimensional gravity. In other words, there is
an {\it infrared anomaly} in the trace sector, and we would expect the effects
of this anomaly to be most important at the largest distance scales. The
consequences of this anomaly for the vacuum energy or cosmological constant
problem may be addressed most directly in the invariant functional integral
approach as well.

The line of reasoning followed in $D=2$ to find the conformal
anomaly and construct the Polyakov-Liouville effective action of two
dimensional quantum gravity may be followed also for $D=4$.
First the general form of the trace anomaly of the energy-momentum
tensor in four dimensions may be calcaluted by heat kernel
regularization of the determinants found by covariant functional
integration over matter fields of arbitrary spin propagating on a
fixed curved manifold. A linear combination of the four invariants, $R^2$,
$R_{ab}R^{ab}$, $R_{abcd}R^{abcd}$, and $\sq R$ is obtained \cite{DesDufIsh}.
Actually, since $\sq R$ is a total divergence in any number of dimensions,
adding it to the effective action gives no contribution to the trace of the
energy momentum tensor. Hence there can be at most three independent local
terms in the general trace anomaly, which therefore may be written in the form
\cite{Duf},
\begin{equation}
T_a^{a \ (matter)} = b\bigl( F + {\textstyle \frac{2}{3}} \sq R
\bigr) + b' G + b'' \ \sq R , \qquad D = 4
\label{anof}
\end{equation}
where
\begin{equation}
F = R_{abcd}R^{abcd} - 2 R_{ab}R^{ab} + {\textstyle {1 \over 3}} R^2
\end{equation}
is the square of the Weyl tensor in four dimensions, and
\begin{equation}
G = R_{abcd}R^{abcd} - 4 R_{ab}R^{ab} + R^2
\end{equation}
is the Gauss-Bonnet integrand. In four dimensions $G$ is a total
divergence, and its integral is a topological invariant, proportional
to the Euler number. Hence we expect its coefficient $b$ to play a
role similar to the $c$ in two dimensions. The coefficients $b$ and
$b'$ have been computed for scalar, Dirac fermion, and vector fields
at one-loop order, with the results \cite{Duf}:
\begin{eqnarray}
b &=& \frac{1}{120 (4 \pi)^2} (N_S + 6 N_F + 12 N_V), \nonumber \\
b'&=& -\frac{1}{360 (4 \pi)^2} (N_S + 11 N_F + 62 N_V).
\end{eqnarray}

If the matter field Lagrangian is invariant under local conformal
(Weyl) transformations classically, then the coefficient $b''$
vanishes at one-loop order. However, in principle it
corresponds to an arbitrary parameter, since $\sq R$ is the variation
of a local action:
\begin{equation}
\sqrt{-g}\ \sq R = -{1 \over 6 }\ g_{ab} {\delta \over \delta g_{ab}} \int d^4
x
\ \sqrt{-g}\ R^2 , \label{varr}
\end{equation}
which receives divergent contributions in general. Thus we allow for
$b'' \neq 0$. In contrast to this term, the first two terms on the
right side of (\ref{anof}) do not arise from local effective actions
in four dimensions. However, in the conformal parameterization of the
metric, the anomaly may be derived from a local effective action, just
as in the two dimensional case \cite{Tsey,AntMot}. In fact, if we
rewrite the trace anomaly (\ref{anof}) in the form,
\begin{eqnarray}
T_a^{a\ (matter)} &=& b F + b' (G - {\textstyle \frac{2}{3}}
\sq R) + [b'' + {\textstyle \frac{2}{3}}(b + b')] \sq R \nonumber \\
&=& {1 \over \sqrt{-g}}{\delta \over \delta
\sigma(x)}S_{anom}[\sigma] ,    \label{anomf}
\end{eqnarray}
the first term is independent of $\sigma$, the second term is linear in
$\sigma$, and are easily integrated, while the $\sigma$ dependence of the
third term is determined by eq. (\ref{varr}). Hence the full anomaly
induced action in four dimensions reads:
\begin{eqnarray}
&&S_{anom}[\bar g,\sigma] = S_{anom}[\bar g]\ b \ \int d^4 x\ \sqrt{-{\bar g}}\
\overline F\ \sigma\ \nonumber\\
&& \ + b' \int d^4 x\ \sqrt{-{\bar g}}\ \Bigl\{ \sigma \bigl[2 \sqb ^2 + 4
\overline R^{ab} {\overline \nabla} _a {\overline \nabla} _b - {\textstyle
\frac{4}{3}} \overline R {\sqb}^2 + {\textstyle \frac{2}{3}} ({\overline
\nabla} ^a \overline R) {\overline \nabla} _a \bigr]\sigma+ \bigl[ \overline G
- {\textstyle \frac{2}{3}} \sqb \overline R \bigr] \sigma
\Bigr\} \nonumber \\
&& \qquad -{\textstyle {1 \over 12}}[b'' + {\textstyle \frac{2}{3}}(b +
b')]\int d^4 x
 \sqrt{-{\bar g}} \bigl[\overline R - 6 \sqb \sigma- 6({\overline \nabla}_a
\sigma)({\overline \nabla}^a \sigma)\bigr]^2 , \label{actan}
\end{eqnarray}
where we have used the expression,
\begin{equation}
R = e^{-2 \sigma}\bigl(\overline R - 6 \sqb \sigma- 6 ({\overline \nabla} _a
\sigma)({\overline \nabla} ^a
\sigma)
\bigr), \label{curv}
\end{equation}
for the Ricci scalar in conformal coordinates. We can again use the
fact that the action must be an invariant function of the total metric
$\bar g e^{2\sigma}$ to write down the fully covariant anomaly-induced
non-local action,
\begin{eqnarray}
S_{anom} &=&-{\textstyle {1 \over 4}} \int d^4 x \sqrt{-g} \ \int d^4 x'
\sqrt{-g'}
\ \bigl[ b F + b' (G - {\textstyle \frac{2}{3}} \sq R) \bigr]_x \nonumber \\
& &\ \bigl[2 \sq ^2 + 4R^{ab} \nabla_a \nabla_b - {\textstyle \frac{4}{3}} R
{\sq}^2 + {\textstyle \frac{2}{3}} (\nabla^a R) \nabla_a \bigr]^{-1}_{xx'}
\bigl[ b F + b' (G - {\textstyle \frac{2}{3}} \sq R) \bigr]_{x'} \nonumber \\
& &\  -{\textstyle \frac{1}{12}}[b'' + {\textstyle \frac{2}{3}}(b + b')] \int
d^4 x \sqrt{-g} R^2.  \label{nonf}
\end{eqnarray}
The non-local term in (\ref{nonf}) is the four dimensional analog of
the non-local form of the $D=2$ Polyakov action, (\ref{loctwo}). As in
two dimensions we must add the anomaly induced action (\ref{actan}) to the
classical Einstein-Hilbert action,
\begin{eqnarray}
S_{cl} &=& {1 \over 2 \kappa} \int d^4 x \sqrt{-g}\ \bigl( R - 2
\Lambda \bigr) \nonumber \\
&=& {1 \over 2 \kappa} \int d^4 x \sqrt{-{\bar g}}\  e^{2 \sigma} \bigl(
 \overline R - 6
\sqb  \sigma- 6 ({\overline \nabla} _a \sigma)({\overline \nabla} ^a \sigma)
\bigr) - {\Lambda \over \kappa} \int d^4 x  \sqrt{-\bar g} \ e^{4 \sigma},
\end{eqnarray}
where $\kappa = 8 \pi G$.

In the case that the fiducial metric is conformally flat, {\it i.e.}
$\bar g_{ab} = e^{2 \bar \sigma}\eta_{ab}$, the effective action for
$\sigma$ may be obtained by a translation, $\sigma\rightarrow \sigma+ \bar
\sigma$ of the following simpler flat space action,
\begin{eqnarray}
& S_{eff}& = S_{anom} + S_{cl}\nonumber \\
& &\ = \int d^4 x \ \Bigl\{ 2b' (\sq \sigma)^2 -[3b'' + 2(b + b')]
\bigl(\sq \sigma+ (\partial_a \sigma)^2 \bigr)^2
+ {3 \over \kappa} e^{2 \sigma} (\partial_a \sigma)^2 - {\Lambda \over
\kappa} e^{4 \sigma} \Bigr\},
\label{flata}
\end{eqnarray}
up to a surface term.

The first important property of this effective action we should note
is that it is {\it completely bounded}, unlike the classical Einstein-Hilbert
action when continued naively to Euclidean signature metrics.

{\it There is no conformal factor problem in the effective theory of the
conformal factor generated by a proper treatment of the measure of the
covariant path integral and trace anomaly in four dimensional gravity.}

In the language of the previous section the constant $C$ of Eq. (\ref{metr})
should be chosen {\it greater} than one half,unlike in the Einstein theory, in
the full effective field theory of the conformal factor generated by $S_{anom}$
in four dimensions, so that the normal continuation to Euclidean signature
applies. Again, the proper derivation of this choice requires a careful
consideration of the canonical formulation of the path integral in phase space
which has not yet been done and remains an open problem. This justification is
important for comparison to the necessarily Euclidean dynamical triangulation
approach as well.

Notice also from the derivation of $S_{anom}$ that although the $\sigma$
independent piece of the gravitational action cannot be determined
from the trace anomaly alone, the $\sigma$ dependence is {\it uniquely
determined} by the general form of the trace anomaly for massless
fields. Thus, whatever else may be involved in the full quantum theory
of gravity in four dimensions at short distance scales, the anomalous
effective action (\ref{actan}) {\it must} be included in the gravitational
action at large distance scales, {\it i.e.} in the far infrared, since the
gravitational effects of massless fields do not decouple and are screened by
nothing at large distances. Graviton ({\it i.e.} spin-two) fluctuations of the
metric should give rise to an effective action of precisely the same form as
$S_{anom}$ with new coefficients $b$ and $b'$, which can be checked at one-loop
order \cite{AMM}. The effective action (\ref{nonf}) or (\ref{flata})
following from the invariant path integral for quantum gravity in the conformal
sector is the starting point for finite volume scaling and infrared
renormalization group comparisons to the dynamical triangulation approach.

\subsection{Finite Volume Scaling and Infrared Renormalization}

An interesting property of the expression (\ref{string}) for the
functional integral over geometries of fixed topology in two
dimensions is that the {\it total} trace anomaly of matter plus ghosts
plus $\sigma$ {\it vanishes} identically. To see this we have only to
realize that the effective action with arbitary coefficient,
\begin{equation}
cS_{anom, Eucl} = + {Q^2 \over 4\pi} \int\,d^2x\, \sqrt {\bar g} \bigl[
({\overline \nabla} \sigma)^2 + \overline R \sigma\bigr] \,
\end{equation}
has an trace anomaly coefficient,
\begin{equation}
c_{\sigma} = 1 + 6 Q^2 \ ,
\end{equation}
where the $1$ is the quantum contribution of a single scalar field
coming from the kinetic term as computed in the last section by the
heat kernel method, and $+6 Q^2$ is the ``classical" contribution
which arises from the simple fact that $S_{anom}$ is not classically
Weyl invariant because of the linear $\bar R\sigma$ term. If we now set
the total anomaly equal to zero
\begin{equation}
c_m + c_{gh} + c_{\sigma} = (N_S + N_F) + (- 26) + (1 + 6Q^2)= 0
\ ,
\end{equation}
we obtain precisely the value of $Q^2 = -c/6$ found in (\ref{central}). The
vanishing of the total trace anomaly coefficient is very important,
since it implies that conformal invariance is restored in the full
quantum theory, when the metric field $e^{2\sigma}\bar g$ is quantized
by the invariant functional integral method. If conformal invariance is
restored then the quantum theory describes a fixed point of the renormalization
group flow, about which we shall say more in the next section. Another
consequence is that all the allowed operators in the theory must have
well-defined
conformal dimension or weight. The conformal weight of an operator $X$
is defined by the behavior of $X$ under a coordinate transformation,
\begin{equation}
\delta_{\xi} X = \xi^a\nabla_a X + (\nabla_a\xi^a)\,{w (X) \over D}\, X \ .
\end{equation}
In classical relativity this just defines the transformation property
of a scalar density of weight $w (X)$. Thus $e^{D\sigma} =
\sqrt{g}$ transforms as a density of weight $D$ in $D$ dimensions,
classically.

In the quantum theory this same transformation is a canonical
transformation generated by the corresponding moment of the energy
momentum tensor $T^{ab}$, {\it viz.}
\begin{equation}
\delta_{\xi} X = i [T_{\xi}, X]\ , \qquad {\rm with}\qquad  T_{\xi} =
\int_{\Sigma}\, d\Sigma_b\, \xi_a\, T^{ab}\ ,
\label{scadim}
\end{equation}
where $\Sigma$ is an arbitrary spacelike Cauchy surface in the space
plus time foliation of the geometry. Because the anomalous action
and its corresponding energy-momentum tensor differ from the classical
ones, the conformal weight $w(X)$ computed this way in the quantum
theory will differ in general from the classical weight. In fact, a
simple computation shows that for a pure exponential operator
\begin{equation}
w (e^{p\sigma}) = p - {p^2 \over 2 Q^2} \ ,
\label{expweight}
\end{equation}
in place of the classical result which is just $p$. However, only
those operators with conformal weight equal to $2$ in $D=2$ dimensions
can be integrated with respect to $d^2 x$ to yield a coordinate
invariant scalar, which may be added to the invariant action. This
implies that the cosmological term in the Polyakov-Liouville action
must be modified to
\begin{equation}
\int d^2x\, e^{2\alpha\sigma} \sqrt {\bar g}\ ,
\end{equation}
where $\alpha$ is determined by requiring the conformal weight to be
$2$:
\begin{equation}
w (e^{2\alpha\sigma}) = 2\alpha - {2\alpha^2 \over Q^2}  = 2\ ,
\end{equation}
or
\begin{equation}
\alpha_{\pm} = {1 \pm \sqrt { 1 - {4\over Q^2}} \over {2 \over Q^2}}\ .
\end{equation}
The negative branch of the square root goes over to the classical
result,
\begin{equation}
\alpha_- \rightarrow 1 \qquad {\rm as} \qquad Q^2 \rightarrow \infty \ .
\end{equation}

As one simple example of the consequences of this anomalous dimension we may
restrict ourselves to the simplest topology, $S^2$ with $h=0$, and
consider the fixed area partition function with $S_{cl} =0$ as well,
\begin{equation}
Z_2(A) \equiv \int J_2\ [{\cal D}\sigma]_{g^{\perp}} \ \exp \bigl( -
cS_{anom} )\ \delta \biggl( \int d^2x\ \sqrt {{-\bar g}}
e^{2\alpha\sigma} - A\biggr)\ .
\end{equation}
By a simple constant shift in $\sigma$,
\begin{equation}
\sigma\rightarrow \sigma+ {\omega \over \alpha}
\label{twoshift}
\end{equation}
we find that $Z(A)$ obeys the scaling relation,
\begin{equation}
Z(A) = e^{-2\omega[Q^2 (1-h)/\alpha + 1]}\, Z(e^{-2\omega} A)
\end{equation}
which implies
\begin{equation}
Z(A) \propto A^{-[Q^2 (1-h)/\alpha +1]} \equiv A^{\Gamma (h) -3} \ ,
\label{twoscale}
\end{equation}
where $\Gamma (h)$ is conventionally called the string susceptibility
for surfaces with genus $h$. For surfaces of genus $0$, {\it i.e.} the
topology of $S^2$, we obtain
\begin{equation}
\Gamma (0) = -Q^2/\alpha + 2 = [c_m-1 -\sqrt {(25-c_m)(1-c_m)}]/12 \ ,
\end{equation}
if the negative sign of the square root is chosen for $\alpha$.
This scaling had been obtained earlier by operator methods in a much
more difficult calculation \cite{KPZ}. By using the properties of the covariant
path integral for $D=2$ dimensional quantum gravity the same result is
almost immediate \cite{DDK}.

Let us now study the consequences of the quantization of the conformal
factor with the four dimensional action (\ref{flata}). The trace of
the energy momentum tensor of this effective scalar theory must
vanish, because of the coordinate invariance of the full theory, for the
same reason as in the $D=2$ case, and as may be explicitly
verified in $D=4$ again. This implies that the beta functions of all
couplings must vanish, but does not preclude the possibility of
nontrivial anomalous scaling dimensions. The
anomalous scaling dimension $\alpha$ in $D=4$ dimensions may be
determined by a calculation only slightly more complicated than the
corresponding one in $2$ dimensions, by requiring that the operator,
$\sqrt{-g} R = -6 e^{2 \alpha \sigma} \bigl( (\partial \sigma)^2 + \sq
\sigma\bigr)$ have conformal weight equal to four. Indeed, direct calculation
of this conformal weight shows that the condition is \cite{AntMot}
\begin{equation}
w(\sqrt{-g} R) = 2 \alpha - {2 \alpha^2 \over Q^2} + 2 = 4.
\end{equation}
This is the necessary condition for the Einstein term to be invariant
under the conformal diffeomorphism symmetry (\ref{scadim}). Equivalently,
one may calculate the logarithmic divergences contributing to the
renormalization of the $\gamma \equiv {3\over \kappa}$ coupling multiplying the
same Einstein term in the classical action and obtain the following beta
function condition:
\begin{equation}
\beta_{\gamma} = \mu {d \over d \mu} \gamma_R = \biggl(2 - 2 \alpha + {2
\alpha^2 \over Q^2}\biggr) \gamma_R = 0.
\end{equation}
The first two terms correspond to the ``classical" scaling dimenension
of the $\gamma$ coupling, for $\alpha \neq 1$, while the $\alpha^2$
term arises from the one-loop counterterm. The vanishing of this beta
function for $\gamma_R \neq 0$ yields a quadratic relation for the
anomalous scaling of $e^{\sigma}$ \cite{AntMot}:
\begin{equation}
1 - \alpha + {\alpha^2 \over Q^2} = 0,
\label{alp}
\end{equation}
which is precisely the same quadratic equation for $\alpha$ as in $2$
dimensions, provided $Q^2$ is defined in $D=4$ by the anomaly
coefficient: of the Gauss-Bonnet term in (\ref{anof}).
\begin{equation}
Q^2 = -32\pi^2 b'\ .
\end{equation}

One observation we can make at this point is that if the expectation
value of the Ricci scalar is different from zero, then the conformal
symmetry must be spontaneously broken. In the semiclassical limit, when
the anomaly induced fluctuations are suppressed by $ 1 \over Q^2$ for
large $Q^2$, $\alpha \rightarrow 1$ and the weight of $R$ is zero, {\it i.e.}
it transforms like a scalar under global conformal transformations. If $\alpha$
differs from unity, then this is no longer the case. This implies that $\langle
R\rangle $ becomes an order parameter for the spontaneous breaking of
global conformal symmetry, in sharp contrast to the classical
situation in which $\langle R\rangle $ can take on any value
consistent with the symmetry. In the classical Einstein equations $R$
is just proportional to the cosmological constant. Because of the
trace anomaly induced by the coordinate invariant (but
Weyl-noninvariant) measure, the cosmological ``constant" problem
reduces to the question of whether conformal symmetry is spontaneously
broken, or restored in the ground state of quantum gravity.

Consider now the correlation function of Ricci scalars, $\langle R(x)
R(x') \rangle$ at two different points. Using the expression,
(\ref{curv}) with $\sigma$ replaced by $\alpha \sigma$, and the free
action which describes the theory at its critical point, we find:
\begin{eqnarray}
\langle R(x) R(x') \rangle &\rightarrow &\overline R^2 \langle e^{-2 \alpha
\sigma(x)}
e^{-2 \alpha \sigma(x')} \rangle \nonumber\\
& \rightarrow &\overline R^2 e^{4 \alpha^2 \langle \sigma (x) \sigma
(x')\rangle}\nonumber \\
&=& \overline R^2 \ \bigg| {H s(x, x')\over 2} \bigg|^{-{4 \alpha^2 \over
Q^2}}\ , \qquad |s(x,x')| \rightarrow \infty ,
\label{cosm}
\end{eqnarray}
where only the dominant infrared behavior has been retained, and $s(x,x')$ is
the invariant distance between the points $x$ and $x'$.

The result (\ref{cosm}) states that the {\it effective} cosmological
``constant" goes to zero with a definite power law behavior for large
distances. In other words, there is screening in the infrared of the
effective value of vacuum energy at larger and larger scales. The
value of the power is {\it universal}, depending only on $Q^2$ which
counts the effective number of massless degrees of freedom. In
particular, it depends neither on the classical value of the
background curvature $\overline R$, nor on the Planck scale. This is essential
for a scale invariant, phenemenologically acceptable solution of the
cosmological constant problem.

These results were obtained in the continuum by treating the
metric $\bar g_{ab}$ as fixed. In other words the transverse,
traceless sector of the theory containing the physical spin-$2$
gravitons was neglected completely. Our basic hypothesis is that these
relations remain true in the infrared when the graviton modes are
included, up to a possible renormalization of the value of $Q^2$. More
precisely we assume that integration over the transverse graviton
modes generates an effective action for $\sigma$ which, when expanded in
powers of derivatives, has the same form as (\ref{actan}) but with
renormalized coefficients. This assumption we have called ``infrared
conformal dominance \cite{AMM}."

Such an assumption is not at all unreasonable from a Wilsonian effective action
point of view. Consider the functional integration over transverse
gravitons (in other words over conformal equivalence classes of
metrics), as well as matter fields, with both infrared and ultraviolet
cut-offs, $\ell$ and $a$. At short distances, graviton effects may
grow uncontrollably due to the presence of the dimensionful Newtonian
coupling $\kappa$, so that $a$ cannot be taken to zero in the
effective action. Conversely, at large distance scales, the
transverse, tracefree fluctuations should be expected to become less
important, so that the effective action should be regular as the
infrared cutoff $\ell$ is removed. If this is the case, then an
infrared stable renormalization group fixed point of the effective low
energy theory is approached as $\ell \rightarrow \infty$. Scale
invariance at this fixed point then requires that the low energy
effective action must be of the form (\ref{actan}) when expanded up to four
derivatives of $\sigma$. Yet, infrared conformal dominance is still at this
point an assumption which can be checked in principle, by finding the
scaling behavior of the invariant functional integral for quantum gravity
in the large $\ell$ or infinite volume limit, as the infrared fixed point is
approached.

In order to derive the scaling behavior of the partition function of
the effective $\sigma$ theory in $D=4$, we require one more anomalous dimension
because of the additional term in the classical action in $D=4$ as
compared with $D=2$. This is the anomalous dimension of the volume
operator in four dimensions, which we denote by $\beta$ (not to be confused
with the $\beta$ function defined previously). The weight of the
pure exponential volume operator is determined by using Eq. (\ref{expweight})
which gives
\begin{equation}
w (\sqrt{\bar g} e^{4 \beta \sigma}) = 4 \beta - {(4 \beta)^2 \over 2 Q^2} = 4
\ ,
\end{equation}
since the normalization of $Q^2$ has been adjusted to give the same relations
in $D=4$ as $D=2$. Hence we find that in order to add the volume or
cosmological
term to the effective action, it must have the anomalous dimension \cite{Sch}
\begin{equation}
\beta_{\pm} = {1 \pm \sqrt { 1 - {8\over Q^2}} \over {4 \over Q^2}}\ .
\end{equation}
where we must again choose the negative sign of the square root in order to
recover the classical result,
\begin{equation}
\beta_- \rightarrow 1 \qquad {\rm as} \qquad Q^2 \rightarrow \infty \ .
\end{equation}
Armed with this information we may now subject the $\sigma$ field to the
analog of the constant shift (\ref{twoshift}) performed in two dimensions,
\begin{equation}
\sigma\rightarrow \sigma+ {\omega \over \beta}
\label{fourshift}
\end{equation}
and use the translational invariance of the invariant integration measure
$[{\cal D} \sigma]$ to find
\begin{equation}
Z(\kappa , \lambda) \equiv \int [{\cal D} \sigma] e^{-S_{eff} [\sigma]} =
e^{- {Q^2 \over \beta} \chi \omega} Z(\kappa e^{- 2 {\alpha\omega\over\beta}},
\lambda e^{4\omega}) \ .
\end{equation}
If we restrict to the fixed topology of $S^4$ for which $\chi= 2$, and define
also the partition function at fixed volume
\begin{equation}
Z(\kappa ; V) \equiv \int [{\cal D} \sigma] e^{\lambda V - S_{eff}
[\sigma]} \ \delta \Bigl( \int d^4 x \sqrt {\bar g} e^{4 \beta \sigma} -
V \Bigr) \ ,
\end{equation}
we obtain
\begin{eqnarray}
Z(\kappa ; V) &=& e^{-2\omega( {Q^2 \over \beta} +2)} Z(\kappa e^{- 2
{\alpha\omega\over\beta}} ; e^{-4\omega} V) \nonumber \\
&=& V^{-({Q^2 \over 2\beta} + 1)} {\tilde Z}(\kappa
V^{-{\frac{\alpha}{2\beta}}})
\nonumber \\ &=& V^{-(4 + Q^2 + Q\sqrt {Q^2-8})/4} {\tilde Z}(\kappa
V^{-{\frac{\alpha}{2\beta}}}) \,
\label{volscal}
\end{eqnarray}
where in the second line we put $e^{4\omega} \propto V$. This analog of the
scaling
relation (\ref{twoscale}) for $D=4$ should be testable by the dynamical
triangulation approach to the functional integration over four geometries which
is
the subject of the last section. Notice also that the limit of very large
volume
$V \rightarrow \infty$ is equivalent to the limit of zero Newtonian constant
$\kappa \rightarrow 0$. This is simply the statement that for scales $\ell >>
L_{Planck} \simeq a$, the ultraviolet cutoff scale becomes {\it irrelevant} and
scaling is determined purely by the infrared fixed point of the theory, the
anomalous dimensions there and the corresponding value of $Q^2$, in complete
accord with the behaviour of critical exponents in the Wilson theory of second
order phase transitions \cite{Wil}. The effective vanishing of the Newtonian
coupling in the large volume limit implies that ordinary perturbation theory
should be applicable
to the calculation of the graviton contribution to the conformal anomaly in the
far infrared \cite {AMM}. The analytic calculation of this contribution and
the complete determination of $Q^2$ and therefore the scaling in
(\ref{volscal})
remains an important unresolved issue, and of course one that should have
consequences for large scale structure in the universe.

\section{The Dynamical Triangulation Definition of the Functional Integral}

A completely different approach to defining the path integral for
reparameterization invariant theories such as quantum gravity is the
dynamical triangulation approach. This is a discretization method that
avoids the introduction of coordinates on manifolds, defining the
functional integration directly over distinct geometries. This
approach has the great advantage of avoiding the gauge fixing problem
of factorizing out the infinite volume of the gauge fiber $\cal G$.
The corresponding disadvantage is that there is no meaning to
reparameterization invariance in the dynamical triangulation approach
and the integration measure is not determined {\it a priori}, or even
necessarily related to the continuum measure constructed in the
previous sections.

Dynamical triangulation is a variation of Regge calculus in which
geometries are constructed by gluing together fundamental simplices of
fixed volume \cite{DT}. In two dimensions the regular simplices are just
equilateral triangles. In four dimensions the four-simplices share
common faces with their neighbors which are $D-1=3$ dimensional
simplices, {\it i.e.} regular tetrahedra of fixed edge length $a$. The
angle between two tetrahedra faces sharing a triangle is
\begin{equation}
\theta = \arccos \Bigl({\textstyle {1\over D}}\Bigr) = 1.3181161
\end{equation}
in $D=4$ dimensions. The volume of a fundamental $J$-simplex is
\begin{equation}
V_J = a^J \Omega_J = {a^J\over J!}\sqrt{{J + 1} \over 2^J},
\label{volj}
\end{equation}
so that the total volume of the simplicial manifold is
\begin{equation}
\int d^4x\ \sqrt{g} \rightarrow N_4 V_4
\label{totvol}
\end{equation}
if $N_4$ is the total number of $4$-simplices in the configuration.

As in the Regge approach, space is regarded as flat inside the $D=4$
simplices with all the curvature residing on the $D-2=2$ dimensional
hinges, {\it i.e.} equilateral triangles with the same fixed edge
length. If $n_i$ is the number of $4$-simplices sharing a given
equilateral triangle $i$, then the deficit angle $\delta_i$ is given
by
\begin{equation}
\delta_i = 2\pi - n_i \theta \ ,
\end{equation}
and the Einstein-Hilbert action takes the value,
\begin{equation}
\int d^4x\ \sqrt{g} \ R \rightarrow \sum_i \delta_i V_2 = (2\pi N_2 -10
\theta N_4) V_2,
\label{Ein}
\end{equation}
where $\sum_i n_i = 10 N_4$ has been used (since each $4$-simplex has
$10$ triangles in its boundary).

Dynamics is now specified by giving an action for each simplicial
triangulation $\cal T$ of the form $S({\cal T}) = \sum_J k_J N_J
({\cal T})$, and summing over all triangulations weighted by this
action with each triangulation otherwise having equal statistical
weight. In this approach there is no diffeomorphism invariance at the
lattice level, and no gauge-fixing problems. As a corollary, neither
is it self evident that the statistical model(s) constructed by the
dynamical triangulation procedure lie in the same universality class
as quantum gravity in four dimensions. Certainly a necessary condition
for that to hold is that the simplicial complexes generated should
approximate a continuous manifold as the lattice becomes finer and
finer. This forces some of the numbers of $J$-subsimplices (for $J =
0, 1, \dots , D$) to satisfy several relations, called
Dehn-Sommerville relations, so that not all of the $N_J$ are
independent \cite{CFL}. In addition we have the Euler relation,
\begin{equation}
\sum_{J=0}^4 (-)^J N_J = \chi \ .
\end{equation}
The net result is that only two of the $N_J$ are independent and the
action may be taken to be of the form,
\begin{equation}
S({\cal T}) = -k_2 N_2 ({\cal T}) + k_4 N_4 ({\cal T}) \ .
\label{actDT}
\end{equation}
Comparing this with the Einstein-Hilbert action, and using the
substitutions (\ref{totvol}) and (\ref{Ein}) leads to the following
identification of the bare simplicial action parameters with the
(unrenormalized) parameters of the continuum theory,
\begin{eqnarray}
k_2 &=& \frac{\sqrt 3 \pi}{4} {a^2 \over \kappa} \nonumber \\
k_4 &=& \frac{5 \sqrt 3}{4} \theta {a^2 \over \kappa} + \frac{\sqrt 5}{96}
\lambda
a^4
\label{param}
\end{eqnarray}
where the numerical coefficients come from eq. (\ref{volj}). In two
dimensions there is no Einstein-Hilbert action if we sum over surfaces
with a fixed topology so that only the total area $N_2$ survives in
the action. One does not need to add any higher derivative couplings to the
action, since these should correspond to irrelevant operators in the infrared,
with
vanishing coefficients at the fixed point of the infrared
renormalization group. Indeed, in the continuum theory, the
coefficients of possible $R^2$ and Weyl-squared terms in the action
vanish at the conformal fixed point \cite{AntMot}. The trace anomaly induced
action
(\ref{actan}) should not be added to the bare lattice action
(\ref{actDT}) either. It is {\it nonlocal} in the full metric and
should be generated {\it dynamically} by the quantum fluctuations of
the simplicial geometries, in analogy with the situation in the two
dimensional case.

Now the partition function or functional integral over four-geometries is
defined in the dynamical triangulation approach as
\begin{equation}
Z_{DT} (k_2 , k_4) \equiv \sum_{{\cal T}} e^{-S({\cal T})} =
\sum_{N_4} Z(k_2 ; N_4) e^{-k_4 N_4} \sim \sum_{N_4} e^{-[k_4
-k_4^c(k_2)] N_4}
\label{partDT}
\end{equation}
where the last step follows if the number of triangulations which can
be made from a given number $N_4$ of $4$-simplices of fixed topology is
exponentially bounded with respect to $N_4$ \cite{Amb,Mig,Abo}. This is
an important condition upon which the entire viability of the dynamical
triangulation approach depends. In four dimensions there is as yet no
rigorous proof of this exponential bound, and further work is necessary.
{\it If} the exponential bound on the number of triangulations with fixed
topology and fixed $N_4$ holds, then the partition function (\ref{partDT})
must exist in a region of the coupling constant plane $k_4 >
k_4^c(k_2)$. By approaching the boundary of this region from above one
can hope to arrive at a continuum limit in which physical correlation
lengths go to infinity when expressed in lattice units. In this
precise sense, one is searching for a ``critical" curve in the $(k_2 ,
k_4)$ plane corresponding to the existence of the a non-trivial
infinite volume limit of the lattice theory defined by (\ref{actDT})
and (\ref{param}). Only on this curve, and as we shall see in a moment,
only at one point on this curve can one hope to find a second order phase
transition where correlation lengths go to infinity in units of the
lattice spacing and the continuum limit of the theory may be defined.

We are interested in the continuum infrared fixed point of the
lattice theory, which can be obtained, in principle, by block
averaging over larger and larger sublattices, thereby determining the
mapping from the bare lattice action with parameters $(k_2, k_4)$ to
the renormalized parameters of the effective action at larger and
larger distance scales. The fixed point of this mapping, if it exists,
is the infrared fixed point of the continuum theory. In practice, in
the dynamical triangulation approach one does not need to perform the
block averaging procedure explicitly, since by taking more and more
four simplices ({\it i. e.} $N_4 \rightarrow \infty$), one is
effectively averaging over larger and larger sublattices in any case.
Provided the exponential bound applies we can write down
the finite volume scaling relation in the dynamical triangulation
approach by simply translating the continuum relation (\ref{volscal})
to the lattice, and obtain
\begin{equation}
Z(k_2 ; N_4)\simeq {\tilde Z}({\tilde k_2}) N_4^{- {Q^2 \over 2 \beta} - 1}
e^{k_4^c (k_2)N_4} \ ,
\label{gamcrit}
\end{equation}
where
\begin{equation}
{\tilde k_2} = k_2 N_4^{\alpha\over 2\beta}\ ,
\label{kscal}
\end{equation}
in the case of fixed $S^4$ topology. The import of this relation for the
lattice simulations is that if $k_2$ is scaled to zero as $N_4 \rightarrow
\infty$, with $\tilde k_2$ fixed, then the fixed volume partition function
$Z(k_2 ; N_4)$
should scale to zero with a definite power of $N_4$ that depends only on
$Q^2$. If $Q^2$ were calculated reliably for gravitons in perturbation theory
there are no free parameters in this prediction. Therefore, a clear test of
this scaling relation is that the power of $N_4$ must be independent of the
parameter ${\tilde k_2}$, if our hypothesis of infrared conformal
dominance is correct.

We are proposing therefore that the procedure for finding the infrared
fixed point of quantum gravity in the continuum limit from the
dynamical triangulation starting point is to first locate the critical
curve in the $(k_2, k_4)$ plane. Then one should attempt to run simulations
with even larger volumes, moving along the critical curve towards the
origin by rescaling $k_2 \rightarrow 0$ as $N_4 \rightarrow \infty$ in
accordance with (\ref{kscal}), keeping $\tilde k_2$ fixed. In this
limit one can then test the validity of the scaling behavior for pure quantum
gravity at its infrared fixed point. Preliminary results of
the simulations with $N_4 \sim 10^4$, seem to indicate a
critical curve $k_4^c (k_2)$ which is approximately linear with a
slope slightly more than $2$\cite{Ambp}. This is encouraging since
the approximate slope of the critical curve should be $5 \theta/\pi = 2.097847$
near the infrared critical point at the origin, according to (\ref{param}),
\begin{equation}
k_4^c (k_2) = {5 \theta \over \pi} k_2 + {\sqrt 5 \over 18\pi}
(\kappa^2\lambda)  (k_2)^2 \ ,
\label{kcrit}
\end{equation}
but more data on bigger simplices are needed to confirm this.

In addition to the scaling of the finite volume partition function,
one may consider a variety of physical observables as well. The
simplest observable with nice scaling behavior is the average
curvature $\langle R\rangle$ defined by \cite{Amb}:
\begin{equation}
\langle R\rangle \equiv \biggl\langle {\int d^4 x{\sqrt g}R\over\int d^4
x{\sqrt g}} \biggr\rangle = {2\pi V_2\over V_4}\biggl[{1\over
N_4}{\partial\over\partial k_2}\ln Z(k_2 ;N_4) -
{5\theta\over\pi}\biggr] \ ,
\end{equation}
where the last equality holds for fixed volume and we used the
relations (\ref{totvol}), (\ref{Ein}) and (\ref{actDT}). Now inserting
the scaling behavior (\ref{gamcrit}) and the expression (\ref{kcrit})
one obtains
\begin{equation}
\langle R\rangle \propto k_2 \propto N_4^{-\alpha/2\beta} \rightarrow 0 \ ,
\label{Rscaling}
\end{equation}
where the proportionality factor depends only on $Q^2$ and the
rescaled $\tilde k_2$. This shows that the average curvature scales to
zero, which is very different behaviour from the classical Einstein
theory where $\langle R\rangle$ is simply proportional to the cosmological
constant. This is a different way of seeing that the {\it effective}
cosmological term vanishes in the infinite volume scaling limit. One question
that has arisen in applying (\ref{Rscaling}) to the available numerical data
with finite volume lattices is whether subtractive normalizations to
(\ref{Rscaling}) are necessary. This is an important point which requires
further elucidation, since without it we do not yet know how to compare
the numerical data directly to the theoretical prediction and interpret the
results.

Similar considerations apply for the curvature-curvature correlator
\begin{eqnarray}
\int d^4 x\ \sqrt{g}\  \langle R(x) R(0) \rangle
& \sim & {1\over N_4} \biggl\langle \Bigl( N_2 - \frac {5}{\pi} \theta N_4
\Bigr)^2 \biggr\rangle \nonumber \\
&=& {1 \over N_4 Z} \Bigl( {\partial \over \partial k_2} -
\frac {5}{\pi} \theta N_4 \Bigr)^2 Z(k_2 ; N_4)\ .
\label{Rcor}
\end{eqnarray}
If we substitute (\ref{kcrit}), (\ref{gamcrit}) and (\ref{kscal}) into
this last expression, a short calculation shows this correlation
function also goes to a finite constant, {\it independent} of $N_4$ in
the large volume limit. This means that the correlator $\langle R(x)
R(0) \rangle$ must fall off faster than $\vert x
\vert^{-4}$ for large $\vert x \vert$, in order for the integral over $x$
to converge. Contrast this convergent behavior of (\ref{Rcor}) in the
conformal invariant phase of 4D gravity to the behavior of the same
quantity in the classical theory. In that case the trace of the
classical Einstein equations, $R=4\Lambda$ rigidly fixes the scalar
curvature in terms of the cosmological constant. Hence (\ref{Rcor})
must {\it diverge} linearly with the volume $N_4$ in the classical
theory, if the cosmological term is non-vanishing.

We conclude that if the quantum fluctuations of the conformal factor
described in this article really dominate in the far infrared, then any
non-zero cosmological term in the continuum theory is screened
completely at the largest distance scales by these fluctuations, or in
other words, the {\it effective} cosmological constant in the scale
invariant vacuum of 4D quantum gravity must vanish.

\section{Conclusions and Open Problems}

In this article the invariant functional integral for coordinate invariant
systems has been constructed and exploited to yield a variety of results,
some previously known and some new. The measure in the coset functional space
of
metrics modulo spacetime diffeomorphisms has been determined by a geometric
construction, and agrees with other methods. The invariant path integral
framework shows clearly the problems with naive semi-classical expansions
around
spaces with normalizable Killing vector fields such as de Sitter space,
where the measure vanishes, and infrared instabilities appear in the
conformal factor of the metric. The effective theory of the conformal factor
in four dimensions constructed also by means of the invariant integral
for gravity provides a novel insight into the importance of quantum
effects in gravitation at large distance scales, whose physical consequences
are still under investigation. The continuum definition of the sum
over geometries provides a useful framework for extracting information about
quantum gravity at its large volume infrared fixed point, independently
of the underlying theory at the Planck scale. This is seen explicitly in
the finite volume scaling relations and comparison to the dynamical
triangulation approach. The application of methods from statistical
mechanics and critical phenomena to qunatum gravity is made possible
by the invariant functional integral formulation.

Many issues remain to be explored further. On a formal level, it is
important to place the construction of the measure on coset spaces
${\cal M} /{\cal G}$ on a more rigourous mathematical foundation. The
connection of this construction to canonical phase space methods has
yet to be fully elucidated. Global issues of topology and coordinate
charts on this very large space have been scarcely touched in this article
or the literature in general. It is important to know if there are global
obstructions to the construction and what the consequences of such
obstructions to physics might be. The Yamabe uniformization problem and
classification of geometries in four dimensions clearly require much more
work as well. Any extension of understanding or exact results for functional
integrals other than Gaussian would definitely be welcome. Investigation of
the simplicial definition of the functional integral is also needed.
Is there indeed an exponential bound to the partition function at fixed
topology and fixed volume? What is the measure in this space of simplices and
does it approximate the continuum measure correctly at large volumes? How
should the finite volume numerical results be compared to the theoretical
predictions which can hold strictly only at the infrared fixed point in the
infinite volume limit?
Analytic  work as well as more efficient algorithms for numerical simulations
on bigger lattices are both necessary in the dynamical triangulation framework.

Of the physics issues raised, a calculation of the graviton contribution
to the infrared anomaly and finite volume scaling is clearly paramount. A
clean test of the scaling relations by the numerical simulations would
be the next most important development, possibly providing a complete solution
of
the long-standing cosmological constant problem \cite{Wei}. Such a dynamical
mechanism based on infrared conformal invariance, internal to the structure of
quantum gravity itself with no additional assumptions is very appealing and
would herald a new era in cosmology and large scale structure formation. The
invariant functional integral over geometries appears to provide us with the
most powerful tool to date for prying open the door at least a bit to a better
understanding of quantum gravity.

\end{document}